\documentclass[10pt]{iopart}

\usepackage{iopams}
\usepackage{graphicx}
\usepackage{cite}

\begin{document}
\title[Power series method for TASEP-based models]{Power series method for solving TASEP-based models of mRNA translation}
\author{S Scott$^1$, J Szavits-Nossan$^1$}
\address{$^1$ SUPA, School of Physics and Astronomy, University of Edinburgh, Peter Guthrie Tait Road, Edinburgh EH9 3FD, United Kingdom}
\ead{jszavits@staffmail.ed.ac.uk}

\begin{abstract}
We develop a method for solving mathematical models of messenger RNA (mRNA) translation based on the totally asymmetric simple exclusion process (TASEP). Our main goal is to demonstrate that the method is versatile and applicable to realistic models of translation. To this end we consider the TASEP with codon-dependent elongation rates, premature termination due to ribosome drop-off and translation reinitiation due to circularisation of the mRNA. We apply the method to the model organism {\it Saccharomyces cerevisiae} under physiological conditions and find excellent agreements with the results of stochastic simulations. Our findings suggest that the common view on translation as being rate-limited by initiation is oversimplistic. Instead we find theoretical evidence for ribosome interference and also theoretical support for the ramp hypothesis which argues that codons at the beginning of genes have slower elongation rates in order to reduce ribosome density and jamming.
\end{abstract}
\submitto{\PB}
\maketitle
\ioptwocol

\section{Introduction}

Translation of a mRNA sequence into a protein is central to normal cell function. How is this process carried out and controlled in the cell is a topic of major interest not only from the standpoint of understanding protein function and regulation, but also for the possibility of making adjustments to the genetic code that would improve yields of foreign and synthetic proteins.

\begin{figure}[hbt]
	\centering\includegraphics[width=8cm]{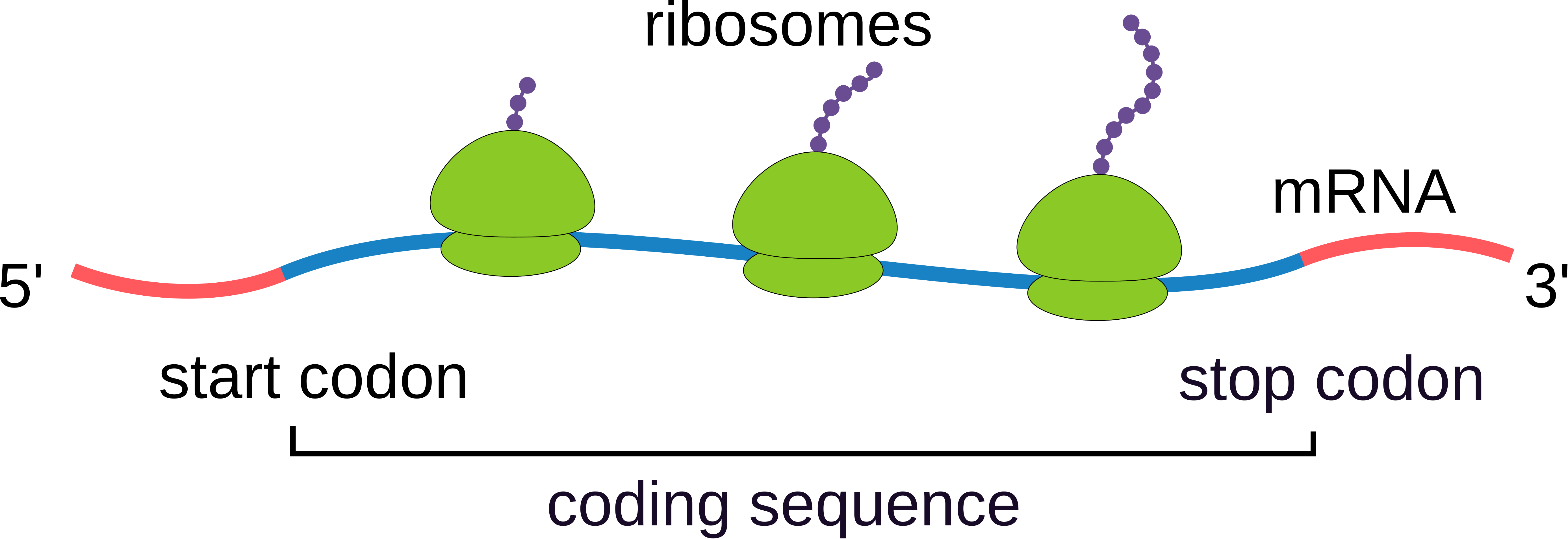}
	\caption{A schematic picture of mRNA translated by ribosomes in the $5'\rightarrow 3'$ direction.}
	\label{fig1}
\end{figure}

Translation is performed by ribosomes that move along the mRNA from the 5' end to the 3' end (Figure \ref{fig1}). The process can be split into three main stages: initiation, elongation and termination. During initiation, the ribosome assembles on a portion of the mRNA before the coding sequence and moves to the start codon where the first amino acid is added to the ribosome. Elongation begins when the ribosome moves to the second codon with a newly amino acid attached to the protein chain. This process is repeated codon by codon until the ribosome encounters the stop codon and detaches itself from the mRNA along with a newly produced protein.

Mathematical modelling of translation has a long history in mathematics, physics and biology. Most of the models that are in use today are based on a model introduced by MacDonald, Gibbs and Pipkin in 1968 \cite{MacDonald68,MacDonald69} and independently by Spitzer in 1970 \cite{Spitzer70}. Spitzer, who was interested in a much broader class of interacting random walks, is also responsible for naming the model the exclusion process due to excluded-volume interactions between the random walkers. The full name of the process relevant to mRNA translation is the totally asymmetric simple exclusion process or TASEP; ``totally asymmetric" means that random walkers (ribosomes) move unidirectionally on a discrete lattice (mRNA) and ``simple" means that they move one lattice site (codon) at a time.

In physics, the TASEP is one of the simplest models belonging to a broad class of {\it driven diffusive systems} \cite{Schmittmann95}. These systems are of great interest because they do not attain thermal equilibrium, even when they settle in the steady state. The question of how to describe nonequilibrium steady states is one of the biggest open questions in statistical physics. For the TASEP in which each random walker occupies one lattice site this problem was solved in full by Derrida, Evans, Hakim and Pasquier \cite{Derrida93} and Sch\"{u}tz and Domany \cite{Schuetz93}, both in 1993. The exact solution described in detail the nature of phase transitions previously discovered by Krug \cite{Krug91}, which sparked a great interest in the model.

Unfortunately, most TASEP-based models which are of interest to modelling translation cannot be solved using techniques developed in Refs. \cite{Derrida93,Schuetz93}. These models account for the correct ribosome length (approximately the length of $10$ codons) \cite{Shaw04}, variable ribosome speed that depends on the codon being translated \cite{Varenne84}, elongation consisting of several intermediate steps \cite{Sorensen89}, nonsensical errors such as premature termination \cite{Gilchrist06,Bonnin17}, translation reinitiation due to mRNA circularisation \cite{Chou03,Gilchrist06,Sharma11,Marshall14} and many more (for a recent review see Ref. \cite{Tuller16}). On the other hand, it is fairly easy to simulate these models on a computer--the main problem is how to interpret the results in terms of the model's parameters. 

A fundamental question in molecular biology is how the mRNA codon sequence affects the translation process and in particular the rate of protein production \cite{Gingold11,Brule17}. In the TASEP the rate of protein production corresponds to the current of ribosomes leaving the stop codon. If we assume that each of 61 codon types\footnote{The remaining three codons are stop codons that do not code for an amino acid.} is translated at a different speed, this leaves us with 61 parameters describing elongation and two parameters describing initiation and termination, and that is only for the basic model. Using stochastic simulations alone in order to understand how these parameters affect the translation process is a difficult, if not a formidable task. A different approach is needed.

In previous work \cite{Nossan18}, Szavits-Nossan, Ciandrini and Romano developed a mathematical method for solving the TASEP with two-step elongation that accounted for tRNA delivery and ribosome translocation \cite{Ciandrini10}. The main idea was to express the steady-state solution as a power series in the translation initiation rate. Using initiation rate as an expansion variable was motivated by the work of Ciandrini, Stansfield and Romano \cite{Ciandrini13}, who inferred initiation rates for {\it Saccharomyces cerevisiae} genes from polysome profiling experiments \cite{Mackay04}. Their study indeed showed that the rate of initiation is the smallest rate in the model for most of the genes.  

In the present study, we apply the power series method to the TASEP that accounts for premature termination due to ribosome drop-off and translation reinitiation due to mRNA circularisation. The main purpose is to show that the method is versatile and practical to use for studying more realistic models of translation. We test the method on the model organism {\it Saccharomyces cerevisiae} and find an excellent agreement with the results of stochastic simulations. 

\section{Methods}
\label{Methods}

\subsection{TASEP-based models of translation}

We model mRNA as one-dimensional lattice consisting of $L$ codons labelled from $1$ (start codon) to $L$ (stop codon) that code for $L-1$ amino acids. We assume that each ribosome occupies $\ell=10$ codons \cite{Ingolia09} and that the ribosome P and A sites are positioned at the fifth and sixth codon respectively, measured from the ribosome's trailing end.

\begin{figure*}[hbt]
	\centering\includegraphics[width=16cm]{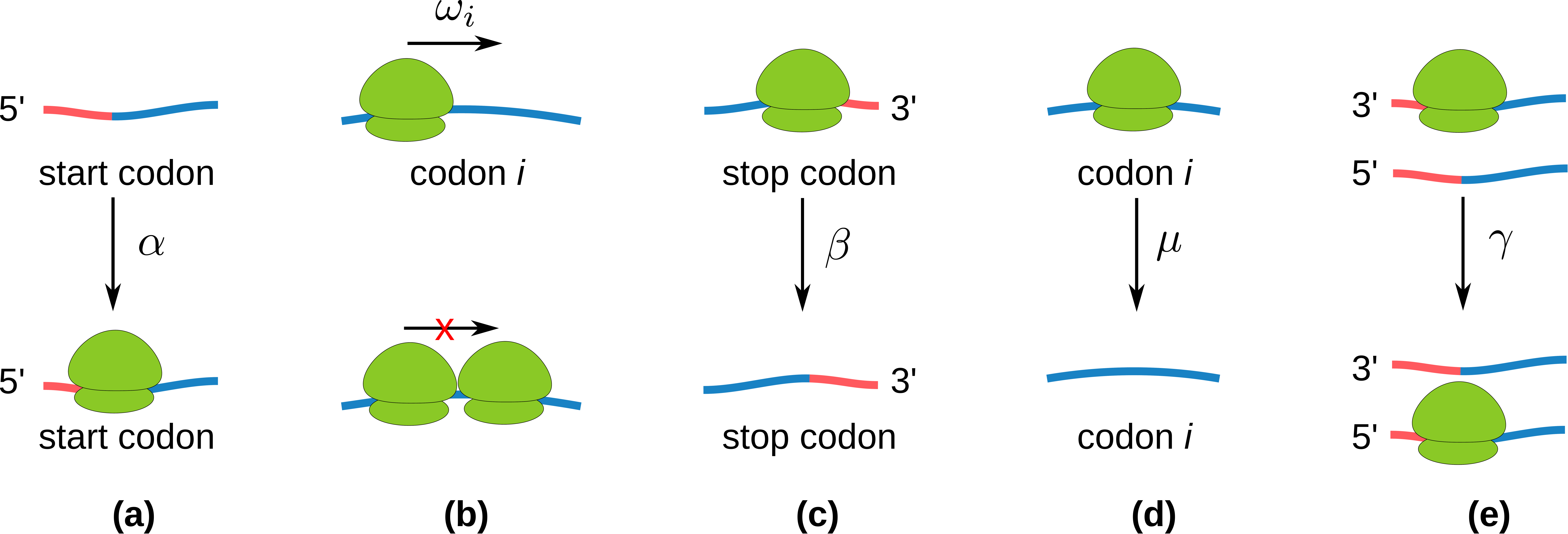}
	\caption{A schematic picture of all the kinetic steps included in the model along with their corresponding rates: (a) initiation (rate $\alpha$), (b) elongation (codon-specific rate $\omega_i$), (c) termination (rate $\beta$), (d) ribosome drop-off (rate $\mu$) and (e) reinitiation (rate $\gamma$).}
	\label{fig2}
\end{figure*}

Translation initiation is a multi-step process which is different in prokaryotic and eukaryotic cells. We model translation initiation as a one-step process occurring at rate $\alpha$ in which a new ribosome is recruited at the start codon so that its P-site and A-site are positioned at the first and second codon, respectively. This one-step process thus encompasses both prokaryotic and eukaryotic translation initiation mechanisms. 

During elongation, a ribosome at codon $i$ receives an amino acid from the corresponding tRNA and translocates to the next codon at rate $\omega_i$, provided there is no ribosome at codon $i+\ell$. Translation terminates once a ribosome A-site reaches the stop codon, releases the polypeptide chain and unbinds from the mRNA at rate $\beta$. For each codon $i=2,\dots,L$ we define the corresponding ribosome occupancy number $\tau_i\in\{0,1\}$,
\begin{equation}
\tau_i=\cases{1 & if codon $i$ is occupied by a ribosome \\ & A-site\\ 0 & otherwise}
\end{equation}
These numbers uniquely determine the configuration of the system which we denote by $C=\{\tau_2,\dots,\tau_{L}\}$. Using this notation, kinetic steps in translation can be summarized as:
\numparts
\begin{eqnarray}
\label{initiation}
&\textrm{(initiation): $\tau_2=0\stackrel{\alpha}{\longrightarrow}1$ if $\tau_2=\dots=\tau_{\ell+1}=0$}\\
\label{elongation}
&\textrm{(elongation): $\tau_i,\tau_{i+1}=1,0\stackrel{\omega_i}{\longrightarrow}0,1$ if $\tau_{i+\ell}=0$}\nonumber\\
&\quad i=2,\dots,L-1\\
\label{termination}
&\textrm{(termination): $\tau_{L}=1\stackrel{\beta}{\longrightarrow}0$.} 
\end{eqnarray}
Equations (\ref{initiation})-(\ref{termination}) constitute the standard model of mRNA translation proposed by MacDonald, Gibbs and Pipkin in 1968 \cite{MacDonald68}. 

In addition to the standard model we also consider premature termination by ribosome drop-off and translation reinitiation due to mRNA circularisation. Ribosome drop-off is a translational error which results in the ribosome being released from the mRNA along with a non-functional polypeptide that is targeted for degradation. We model ribosome drop-off as a one-step process in which a ribosome at codon $i=2,\dots,L-1$ unbinds from the mRNA at rate $\mu$,
\begin{equation}
\label{drop-off}
\textrm{(ribosome drop-off): $\tau_i=1\stackrel{\mu}{\longrightarrow}0$}
\end{equation}
for $i=2,\dots,L-1$. Translation reinitiation is a mechanism by which the ribosome that just finished translation may pass directly from the 3' end to the 5' and initiate another round of translation (see \cite{Marshall14} and references therein). This is made possible by interactions between the two ends of the mRNA resulting in a mRNA circularisation \cite{Wells98}. For translation reinitiation we consider the simplest one-step process in which a ribosome recognizes the stop codon, releases the polypeptide chain and reinitiates translation at rate $\gamma$,
\begin{eqnarray}
\label{reinitiation}
&\textrm{(translation reinitiation): $\tau_2,\tau_{L}=0,1\stackrel{\gamma}{\longrightarrow}1,0 $}\nonumber\\ &\qquad\textrm{if $\tau_{2}=\dots\tau_{\ell+1}=0$}.
\end{eqnarray}
\endnumparts
A schematic picture of the steps (\ref{initiation})-(\ref{reinitiation}) is presented in Fig. \ref{fig2}. There are other mechanisms that we do not consider here. For example, two-step elongation consisting of tRNA delivery to the ribosome A-site followed by translocation has been previously analyzed in Ref.~\cite{Nossan18}.

\subsection{Ribosome current and density}

Our goal is to compute the rate of protein synthesis $J$ and ribosome (A-site) density $\rho_i$. The rate of protein synthesis $J$ is equal to the total current of ribosomes leaving the stop codon,
\begin{equation}
\label{sythesis-rate}
J=\beta\langle\tau_{L}\rangle+\gamma\left\langle\tau_L\prod_{i=2}^{\ell+1}(1-\tau_i)\right\rangle.
\end{equation}
Here the first term is due to termination and the second term is due to translation reinitiation. Th current $J$ is not conserved across the coding mRNA (unless we ignore premature termination) and is different from the current of ribosomes initiating translation 
\begin{equation}
\label{initial_current}
J_{\textrm{in}}=\alpha\left\langle\prod_{i=2}^{\ell+1}(1-\tau_i)\right\rangle+\gamma\left\langle\tau_L\prod_{i=2}^{\ell+1}(1-\tau_i)\right\rangle.
\end{equation}
For the rest of the codons the ribosome current (number of ribosomes moving from codon $i$ to codon $i+1$ per second) is given by
\begin{equation}
\label{current_i}
J_i=\omega_i\left\langle\tau_i\prod_{j=i+1}^{i+\ell}(1-\tau_j)\right\rangle,\; i=2,\dots,L-1.
\end{equation}

Other important observables are the ribosome (A-site) density $\rho_i$ at codon $i$ and the average density $\rho$ defined as 
\begin{eqnarray}
\label{ribosome-density}
\rho_i=\langle\tau_i\rangle,\\
\label{total-density}
\rho=\frac{1}{L-1}\sum_{i=2}^{L}\rho_i.
\end{eqnarray}

The averaging $\langle\dots\rangle$ in Eqs. (\ref{sythesis-rate})-(\ref{total-density}) is taken with respect to the steady-state probability $P(C)$ to find the system in a configuration $C$,
\begin{eqnarray}
\langle\dots\rangle&=&\sum_{C}(\dots)P(C)=\\
&=&\sum_{\tau_2=0,1}\dots\sum_{\tau_{L+1}}(\dots)P(\tau_2,\dots,\tau_{L+1}).
\end{eqnarray}
The steady-state probability $P(C)$ satisfies a master equation,
\begin{equation}
\label{master-equation}
0=\sum_{C'}W(C'\rightarrow C)P(C')-\sum_{C'}W(C\rightarrow C')P(C),
\end{equation}
where $W(C\rightarrow C')$ denotes the rate of transition from configuration $C=\{\tau_2,\dots,\tau_{L}\}$ to $C'=\{\tau'_{2},\dots,\tau'_{L}\}$.

\subsection{Model parameters}

In this paper we study {\it S. cerevisiae} as a model organism using model parameters presented in Table \ref{table1}. 

\begin{table}
\caption{\label{table1}List of TASEP parameters for {\it S. cerevisiae}.}
\begin{indented}
\item[]\begin{tabular}{@{}llll}
\br
Parameter & Variable & Value & Reference\\
\mr
number of codons & $L$ & 25--4093 & Ref. \cite{}\\
ribosome size & $\ell$ & $10$ codons & Ref. \cite{Ingolia09}\\
initiation rate & $\alpha$ & 0.005--4 s$^{-1}$ & Ref. \cite{Ciandrini13}\\
elongation rate & $\omega_i$ & 1--16 s$^{-1}$ & Ref. \cite{Ciandrini13}\\
termination rate & $\beta$ & $35$ s$^{-1}$ & - \\
drop-off rate & $\mu$ & $1.4\cdot 10^{-3}$ s$^{-1}$ & Ref. \cite{Sin16}\\
reinitiation rate & $\gamma$ & - & -\\
reinitiation efficiency & $\eta$ & 0--1 & -\\
\br
\end{tabular}
\end{indented}
\end{table}

Translation initiation rates were obtained in Ref. \cite{Ciandrini13} by matching a theoretical prediction for the total density to the density obtained from polysome profiling experiments \cite{Mackay04}. We note that the TASEP-based model used to estimate initiation rates in Ref. \cite{Ciandrini13} is different from the TASEP-based models we consider here. Because our main goal here is to assess the applicability of the power series method, we use the same values for initiation rates as in Ref. \cite{Ciandrini13}, but note that these may be different from the true (physiological) values. Codon-specific translation elongation rates $\omega_i$ were computed according to
\begin{equation}
    \omega_i=\frac{k_i r_{\textrm{trans}}}{k_i+r_{\textrm{trans}}},
    \label{omega-i}
\end{equation}
where $k_i$ is the tRNA delivery rate for the amino acid corresponding to codon $i$ and $r_{\textrm{trans}}=35$ codons/s is the rate of ribosome translocation \cite{Savelsbergh03}. The values of $k_i$ are assumed to be proportional to tRNA gene copy numbers and were taken from Ref. \cite{Ciandrini13}. The rate of termination is assumed to be large and not limiting for translation; for that purpose we set $\beta=\gamma=35$ s$^{-1}$. The rate of ribosome drop-off is assumed to be the same as for {\it E. coli}, whose value was estimated at $1.4\cdot 10^{-3}$ s$^{-1}$ in Ref. \cite{Sin16}. We are not aware of any estimates of the reinitiation rate $\gamma$ in the literature. Instead we introduce a new parameter $0\leq\eta\leq 1$ that we call reinitiation efficiency,
\begin{equation}
    \eta=\frac{\gamma}{\gamma+\beta},\quad \gamma=\frac{\eta\beta}{1-\eta}
\end{equation}
which measures the value of $\gamma$ relative to the total termination rate $\gamma+\beta$. For example, $\eta=0$ and $\eta=1$ correspond to $\gamma=0$ and $\gamma\rightarrow\infty$, respectively. 

\begin{figure}[hbt]
\centering\includegraphics[width=8cm]{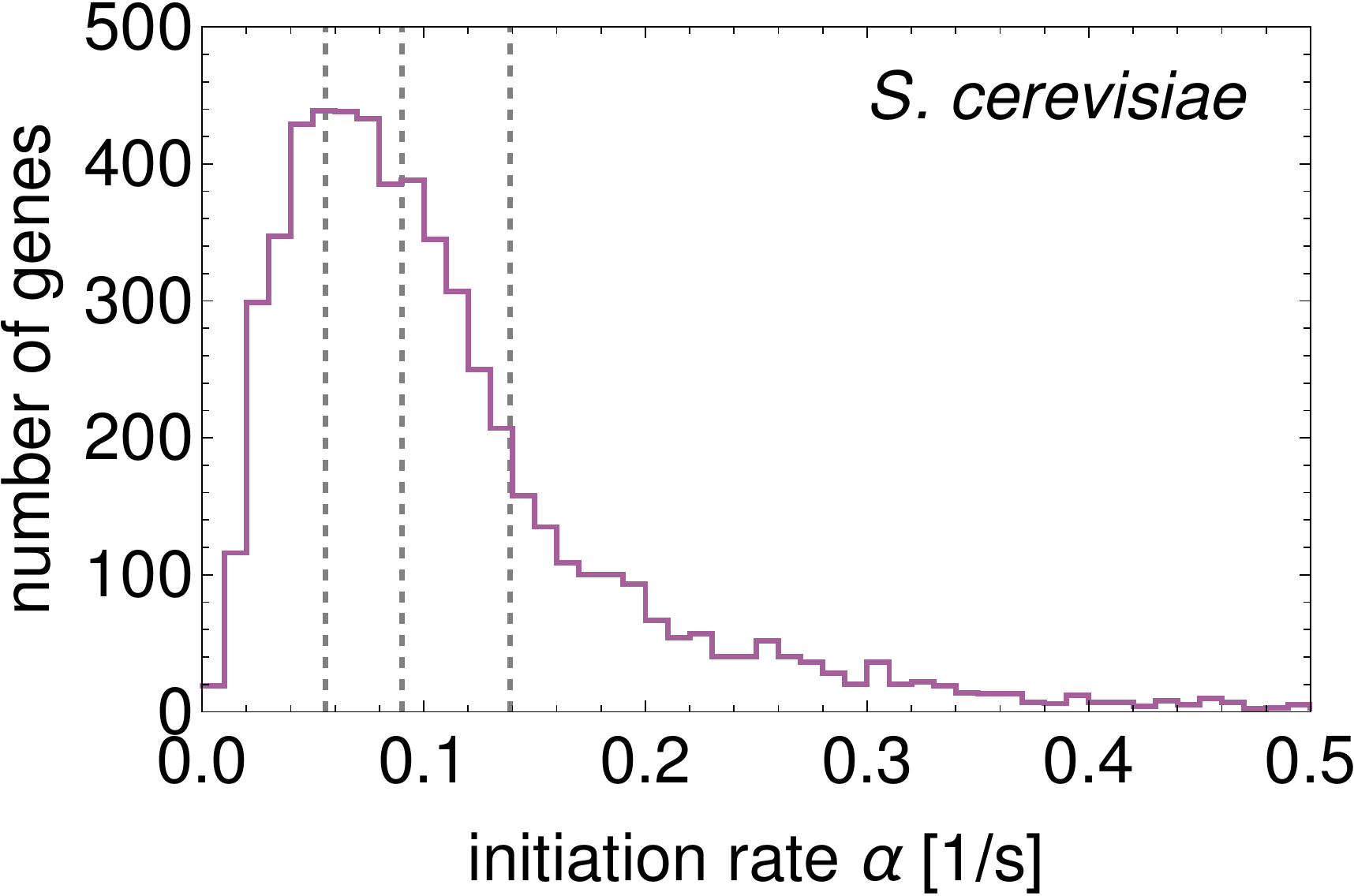}
\caption{Distribution of translation initiation rates for the {\it S. cerevisiae} genome taken from Ref. \cite{Ciandrini13}. Vertical dashed lines are quartile values $0.05578$, $0.09037$ and $0.13889$.}
\label{fig3}
\end{figure}

\subsection{Power series method}

The power series method, previously developed in Refs. \cite{Nossan13,Nossan18}, represents $P(C)$ as a power series in the translation initiation rate $\alpha$,
\begin{equation}
\label{power-series}
P(C)=\sum_{n=0}^{\infty}c_n(C)\alpha^n.
\end{equation}
Here $c_n(C)$ are unknown coefficients that depend on configuration $C$ and other rates. From the fact that all $P(C)$ must sum to $1$, we immediately get that
\begin{equation}
\label{sum-c_n}
\sum_{C}c_n(C)=\cases{1, & $n=0$\\ 0 & $n\geq 1$.}
\end{equation}
While it is possible to expand $P(C)$ in other rates, we expect translation initiation rate to be much smaller that any other rate. Indeed, the median value of $\alpha$ estimated for the {\it S. cerevisiae} genome is an order of magnitude smaller than any of the elongation rates \cite{Ciandrini13}. That allows us to approximate series expansion of $P(C)$ by the first $K$ terms (\ref{power-series})
\begin{equation}
\label{taylor-polynomial}
P(C)\approx c_0(C)+c_1(C)\alpha+\dots+c_K(C)\alpha^K.
\end{equation}
It needs to be emphasized that keeping only a finite number of terms may lead to significant errors when the rate of initiation is high. This in turn may lead to non-physical values of $P(C)<0$ or $P(C)>1$. Of course if that happens the method is not applicable for that choice of $\alpha$ and one has to compute higher-order terms.

In order to find $c_n(C)$, we insert the power series (\ref{power-series}) back into the master equation (\ref{master-equation}) and collect all the terms that contain $\alpha^n$. These terms must all sum to zero because the left hand side of the stationary master equation (\ref{master-equation}) is equal to zero.  Before we write down a general expression for $c_n(C)$ we need to distinguish between $W(C\rightarrow C')=\alpha$ and $W(C\rightarrow C')\neq\alpha$. For that purpose we introduce an indicator function $I_{C,C'}$ defined as
\begin{equation}
    I_{C,C'}=\cases{1 & $C\rightarrow C'$ is an initiation event \\
    0 & otherwise.}
\end{equation}
This allows us to write $W(C\rightarrow C')$ as 
\begin{eqnarray}
    W(C\rightarrow C')&=&\alpha I_{C,C'}+W(C\rightarrow C')(1-I_{C,C'})\nonumber\\
    &=& \alpha I_{C,C'}+W_0(C\rightarrow C')
\end{eqnarray}
where $W_0(C\rightarrow C')=(1-I_{C,C'})W(C\rightarrow C')$. Inserting $P(C)$ from (\ref{power-series}) into (\ref{master-equation}) and equating the sum of all terms containing $\alpha^n$ to 0 gives the following equation for $c_n(C)$ for $C\neq \emptyset$
\begin{eqnarray}
c_n(C)&=&\frac{1}{e(C)}\left(\sum_{C'}W_0(C'\rightarrow C)c_n(C')\right.\nonumber\\
&+&\left.\sum_{C'}c_{n-1}(C')I_{C',C}-c_{n-1}(C)\sum_{C'}I_{C,C'}\right),\label{cn-eq}
\end{eqnarray}
where $e(C)$ is the total exit rate from $C$ excluding initiation 
\begin{equation}
    e(C)=\sum_{C'}W_0(C\rightarrow C').
\end{equation}
For $C=\emptyset$ we can use Eq. (\ref{sum-c_n}) instead which gives
\begin{equation}
    c_n(\emptyset)=\delta_{n,0}-\sum_{C'\neq\emptyset}c_n(C').
\end{equation}
The equation (\ref{cn-eq}) applies to $n\geq 1$. For $n=0$ the equation is simpler and reads
\begin{equation}
e(C)c_0(C)=\sum_{C'}W_0(C'\rightarrow C)c_0(C')\label{c0-eq}
\end{equation}
Notice that (\ref{c0-eq}) is the same as the original master equation in which the rate of initiation is set to $0$. If there is no initiation then $c_0(C)=1$ if $C=\emptyset$ and is $0$ otherwise,
\begin{equation}
\label{c0}
c_0(C)=\cases{1, & $C=\emptyset$\\ 0, & otherwise.}
\end{equation}
The power series method can be understood as a perturbation theory in which translation initiation events can be seen as a small ``perturbation'' of the empty lattice. 

An important consequence of (\ref{c0}) is that any $c_n(C)$ for which the index $n$ is smaller than the total number of ribosomes $N(C)$ in $C$ is equal to zero, or alternatively
\begin{equation}
\label{cn-non-zero}
c_n(C)\neq 0\textrm{ only if }n\geq N(C)=\sum_{i=2}^{L}\tau_i.
\end{equation}
This result is not obvious but follows from the Markov chain tree theorem \cite{Chaiken78} (also known as Schnakenberg network theory in physics \cite{Schnakenberg76}). We refer the reader to Ref.~\cite{Nossan18b} in which we proved (\ref{cn-non-zero}) for the standard TASEP with particles of size $\ell=1$, but the same arguments pertain to the models studied in this paper.

The result in (\ref{cn-non-zero}) simplifies the calculation of $c_n(C)$ considerably. For $n=1$, we only have to consider configurations with one ribosome ($C=1_i$ for $i=2,\dots,L$) or less ($C=\emptyset$). For $n=2$, only configurations with two ribosomes ($C=1_i 1_j$, $i=2,\dots,L-\ell$, $j=i+\ell,\dots,L$) or less ($C=1_i$ for $i=2,\dots,L$ and $C=\emptyset$) need to be studied and so on. This simplification is central to the success of the power series method, allowing us to solve many TASEP-based models for which no exact solution is known. 

\subsubsection{First-order approximation}

According to (\ref{cn-non-zero}) we can ignore all configurations with more than one ribosome. Using (\ref{cn-eq}) we get
\numparts
\begin{eqnarray}
\label{c1-2-eq}
&c_1(1_2)=\frac{1}{\omega_2+\mu}+\frac{\gamma}{\omega_2+\mu}c_1(1_{L})\\
\label{c1-i-eq}
&c_1(1_i)=\frac{\omega_{i-1}}{\omega_i+\mu}c_1(1_{i-1}),\;i=3,\dots,L-1\\
\label{c1-L-eq}
&c_1(1_{L})=\frac{\omega_{L-1}}{\beta+\gamma}c_{L-1}(1_{L-1})\\
\label{c1-0-eq}
&c_0(\emptyset)=\sum_{i=2}^{L-1}\mu c_1(1_i)+\beta c_1(1_{L}).
\end{eqnarray}
\endnumparts
Here we adopted a shorter notation in which $1_i$ denotes a configuration with ribosome at codon $i$, and the rest of the mRNA is empty. First we solve equations (\ref{c1-i-eq}) and (\ref{c1-L-eq}) recursively yielding coefficients $c_1(1_i)$ for $i=3,\dots,L$ that depend on $c_1(1_2)$. After that we insert $c_1(1_{L})$ back into equation (\ref{c1-2-eq}) and find $c_1(1_2)$. Once we have found $c_1(1_2)$ we solve the rest of the equations recursively. Altogether the solution is
\numparts
\begin{eqnarray}
\label{c1-i-solution}
&c_1(1_i)=\frac{\prod_{j=2}^{i}\frac{\omega_j}{\omega_j+\mu}}{\omega_i\left(1-\frac{\gamma}{\beta+\gamma}\prod_{j=2}^{L-1}\frac{\omega_j}{\omega_j+\mu}\right)},2\leq i\leq L-1\\
\label{c1-L+1-solution}
&c_1(1_{L})=\frac{\prod_{j=2}^{L-1}\frac{\omega_j}{\omega_j+\mu}}{(\beta+\gamma)\left(1-\frac{\gamma}{\beta+\gamma}\prod_{j=2}^{L-1}\frac{\omega_j}{\omega_j+\mu}\right)}\\
\label{c1-empty-solution}
&c_1(\emptyset)=-\sum_{i=2}^{L}c_1(1_i).
\end{eqnarray}
\endnumparts
In the last expression we used the property in (\ref{sum-c_n}) which says that all first-order coefficients must sum to zero.

\subsubsection{Second-order approximation}

For the second order, $c_2(C)\neq 0$ only if $C$ contains at most two particles. The equations for $c_2(C)$ are more complicated than for $c_1(C)$ and must be solved numerically. 

Before we write the equations, we first introduce Kronecker delta function $\delta_{ij}$ and unit step function $\theta[i]$ defined as
\begin{equation}
    \delta_{ij}=\cases{1 & $i=j$\\0 & $i\neq j$}\quad \theta[i]=\cases{1 & $i\geq 0$\\ 0 & $i<0$}.
\end{equation}
These two functions allows us to write the equations for $c_2(C)$ in a compact form which reads
\begin{eqnarray}
\label{c2-ij}
c_2(1_i 1_j)&=&\frac{\delta_{i,2}}{e(1_i 1_j)}c_1(1_j)+\frac{\theta[i-3]\omega_{i-1}}{e(1_i 1_j)}c_2(1_{i-1}1_j)\nonumber\\
&&+\frac{\theta[j-i-\ell-1]\omega_{j-1}}{e(1_i 1_j)}c_2(1_i 1_{j-1})\nonumber\\
&&+\frac{\delta_{i,2}\theta[L-\ell-j]\gamma}{e(1_i 1_j)}c_2(1_j 1_L),
\end{eqnarray}
where $e(1_i 1_j)$ is the total exit rate from configuration $1_i 1_j$ excluding initiation,
\begin{eqnarray}
e(1_i 1_j)&=&\theta[j-i-\ell-1]\omega_i+(1-\delta_{j,L})\omega_j+\delta_{j,L}\beta\nonumber\\
&&+\theta[i-\ell-2]\delta_{j,L}\gamma+2\mu.
\end{eqnarray}
Without reinitiation ($\gamma=0$), $c_2(1_i 1_j)$ depends only on $c_2(1_{i-1} 1_j)$ and $c_2(1_i 1_{j-1})$, except for $i=2$ for which it also depends on the known coefficient $c_1(1_j)$. The equation (\ref{c2-ij}) for $\gamma=0$ can be thus solved recursively starting from $i=2$ and $j=2+\ell$, for which $c_2(1_2 1_{\ell+2})=c_1(1_{\ell+2})/(\omega_{\ell+2}+2\mu)$, and iterating over $i=2,\dots,L-\ell$ and $i+\ell\leq j\leq L$.

This procedure cannot be immediately applied to the model with reinitiation (in which $\gamma>0$), because $c_2(1_2 1_j)$ also depends on $c_2(1_j 1_L)$ for $\ell+2\leq j\leq L-\ell$. Instead, the idea is to find coefficients $c_2(1_j 1_L)$ independently and insert them back into Eq. (\ref{c2-ij}), which can be then solved as before. 

To this end, we start from $i=2$ and $j=\ell+2$ in which case $c_2(1_2 1_{\ell+2})$ is a linear combination of $c_1(1_{\ell+2})$ and $c_2(1_{\ell+2} 1_L)$,
\begin{eqnarray}
c_2(1_2 1_{\ell+2})&=&\frac{1}{e(1_2 1_{\ell+2})}c_1(1_{\ell+2})\nonumber\\
&+&\frac{\gamma}{e(1_2 1_{\ell+2})}c_2(1_2 1_{\ell+2})
\end{eqnarray}
Next, we iterate Eq. (\ref{c2-ij}) over $\ell+3\leq j\leq L$ for fixed $i=2$, which can be done explicitly yielding 
\begin{equation}
c_2(1_2 1_j)=\sum_{m=\ell+2}^{j}\left[F_{2,j}^{(m)}c_2(1_m 1_L)+G_{2,j}^{(m)}c_1(1_m)\right],
\label{c2-2j-linear-combination}
\end{equation}
where $F_{2,j}^{(m)}$ and $G_{2,j}^{(m)}$ are given by
\numparts
\begin{eqnarray}
&& F_{2,j}^{(m)}=\gamma\theta[L-\ell-m]G_{2,j}^{(m)},\\
&& G_{2,j}^{(\ell+2)}=\frac{1}{e(1_2 1_{\ell+2})}\prod_{k=\ell+2}^{j-1}B_k\\
&& G_{2,j}^{(m)}=\frac{1}{e(1_2 1_{m})}\frac{\prod_{k=\ell+2}^{j-1}B_k}{\prod_{k=\ell+2}^{m-1}B_k},\;m=\ell+3,\dots,L,
\end{eqnarray}
\endnumparts
and $B_k=\omega_k/e(1_2 1_{k+1})$. If we now choose $j=L$ we get what we were looking for -- an equation that contains coefficients $c_2(1_m 1_L)$ and $c_1(1_m)$. We can now repeat this procedure for $i=3$ by iterating over $j$ until we get the equation for $c_2(1_3 1_L)$, which will again contain $c_2(1_m 1_L)$ and $c_1(1_m)$ and so on. At the end of this procedure we will have a linear system of $L-\ell-1$ equations for $L-\ell-1$ coefficients $c_2(1_2 1_L),\dots,c_2(1_{L-\ell}1_L)$ that can be solved numerically using standard techniques. Once these coefficients are computed, we can then proceed to iterate Eq. (\ref{c2-ij}) as we did before for the model without reinitiation.

Once all two-particle second order coefficients are computed, we can easily compute the remaining one-particle coefficients $c_2(1_i)$ from the following equations,
\numparts
\begin{eqnarray}
\label{c2-2-eq}
c_2(1_2)&=&\frac{1}{\omega_2+\mu}c_1(\emptyset)+\beta c_2(1_2 1_L)+\frac{\gamma}{\omega_2+\mu}c_2(1_{L})\nonumber\\
&+&\mu\sum_{j=\ell+2}^{L-1}c_2(1_2 1_j)\\
\label{c2-i-eq}
c_2(1_i)&=&\frac{\omega_{i-1}}{\omega_i+\mu}c_2(1_{i-1})+\theta[L-\ell-i]\beta c_2(1_i 1_L)\nonumber\\
&+&\mu\sum_{j=2}^{i-\ell}c_2(1_j 1_i)+\mu\sum_{j=i+\ell}^{L-1}c_2(1_i 1_j)\nonumber\\
&-&\theta[i-\ell-2]c_1(1_i),\;i=3,\dots,L-1\\
\label{c2-L-eq}
c_2(1_{L})&=&\frac{\omega_{L-1}}{\beta+\gamma}c_{L-1}(1_{L-1})-c_1(1_L)\nonumber\\
&+&\mu\sum_{j=2}^{L-\ell}c_2(1_j 1_L).
\end{eqnarray}
\endnumparts
Finally, we can compute $c_2(\emptyset)$ using Eq. (\ref{sum-c_n}), which completes the procedure of finding all second-order coefficients $c_2(C)$.

\subsubsection{Higher-order approximations.} In principle, we can use Eq. (\ref{cn-eq}) to compute $c_n(C)$ for any order $n$. In practice, we are limited by the amount of computer memory we need for storing these coefficients. In the model with translation reinitiation, we are further limited by the size of the linear system that can be solved numerically. In the present work we computed ribosome density up to the fourth order in the model without reinitiation and up to the second order in the model with reinitiation.

\subsection{Monte Carlo simulations}

All Monte Carlo simulation were performed using the Gillespie algorithm. In the first part of the simulation we checked the total density $\rho$ every $100\cdot L$ updates until the percentage error between two values of the total density $\rho$ was less than $0.1\%$. After that we ran the simulation for further $M=10^4\cdot L$ updates during which we computed the time average of $\rho_i$ defined as
\begin{equation}
    \rho_{i}=\frac{1}{T}\sum_{k=1}^{M}\tau_{i}^{(k)}\Delta t^{(k+1)},
\end{equation}
where $\tau_{i}^{(k)}$ is the value of $\tau_i$ ($1$ if codon $i$ is occupied by the ribosome's A-site and $0$ otherwise) at $k$-th update in the simulation, $\Delta t^{(k)}=t^{(k)}-t^{(k-1)}$, $t^{(k)}$ is the time of the $k$-th update, $t^{(0)}=0$ and $T=t^{(M)}$.
\begin{figure*}[htb]
\centering
\includegraphics[width=8cm]{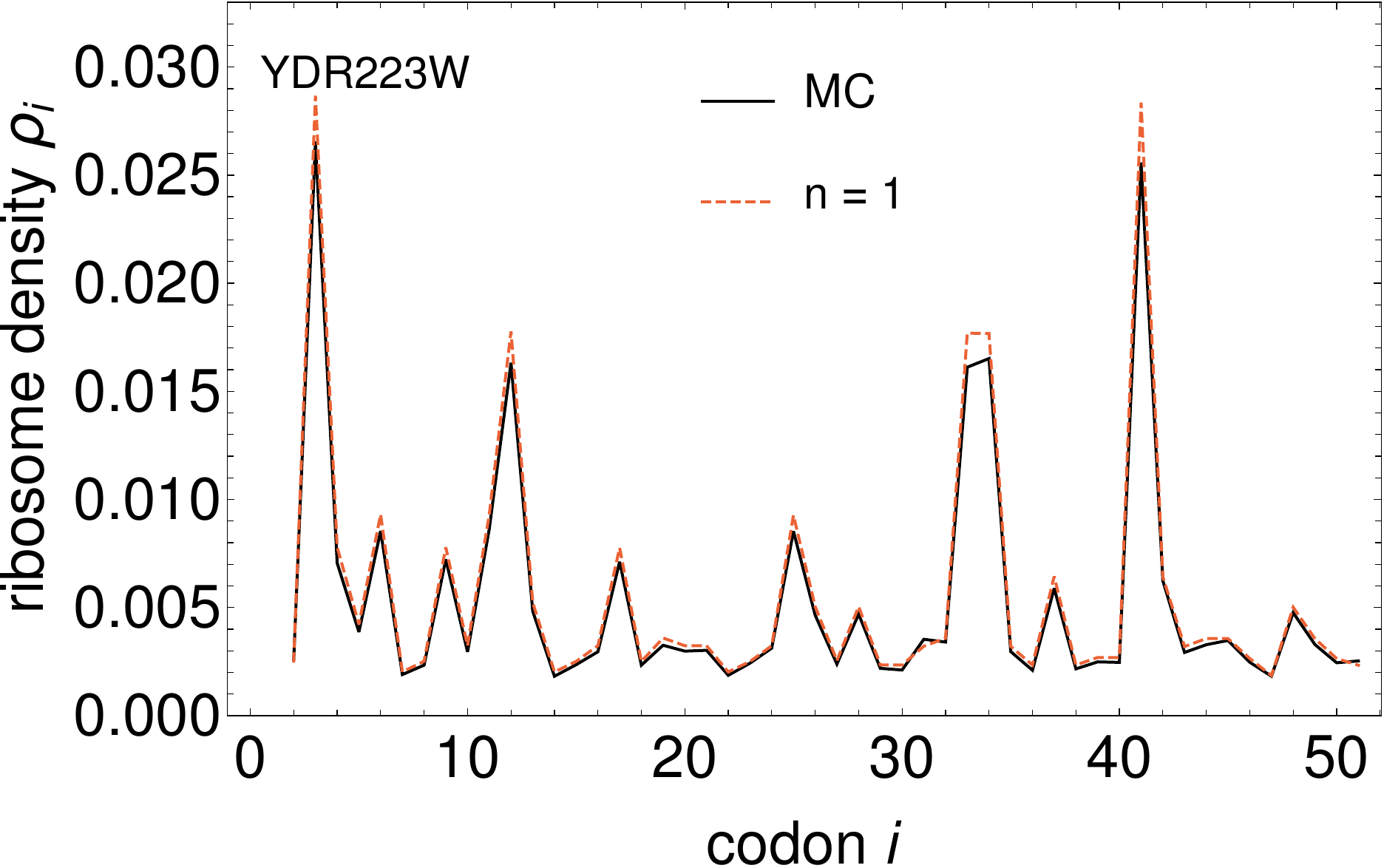}
\includegraphics[width=8cm]{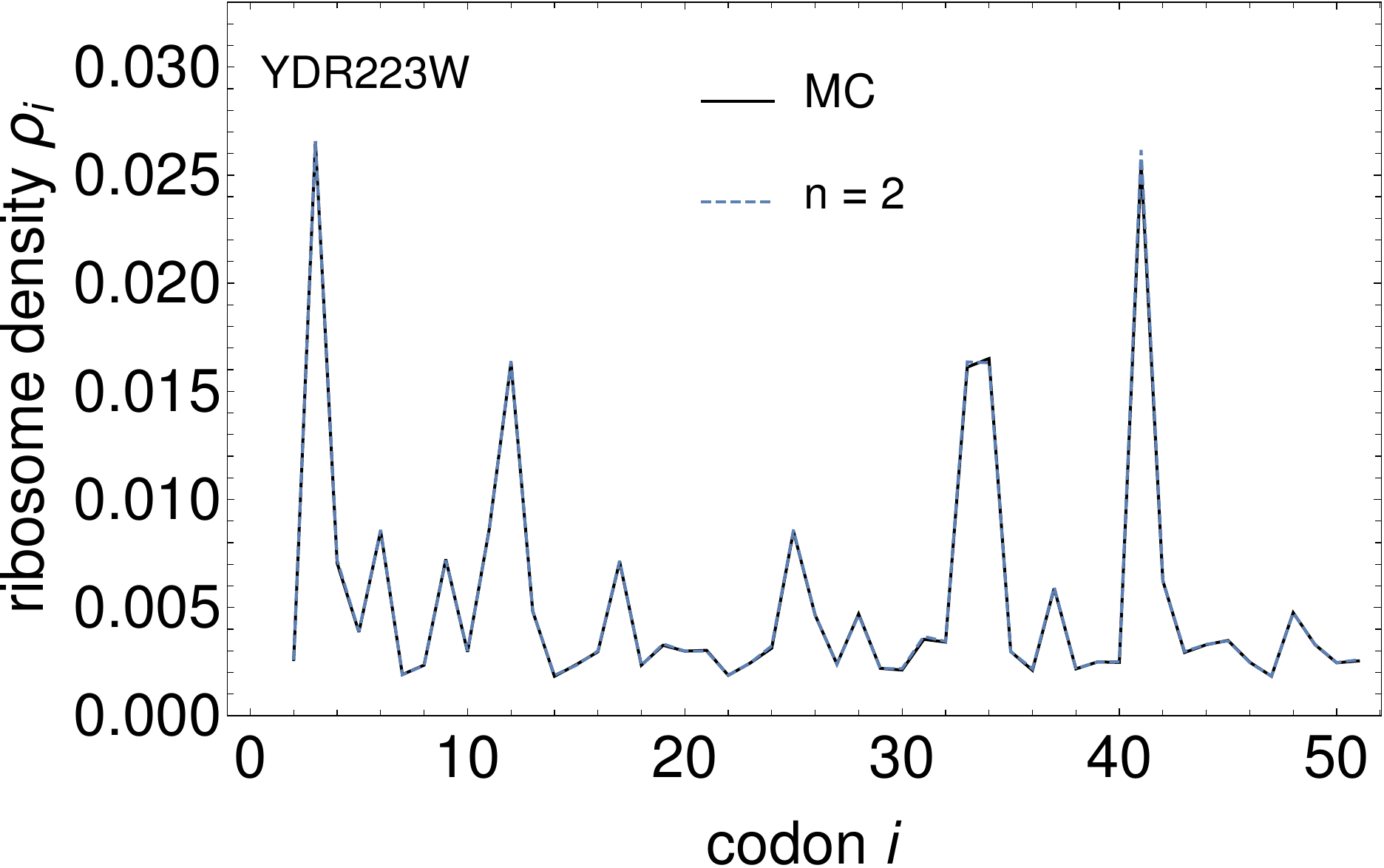}\\
\vspace{1em}
\includegraphics[width=8cm]{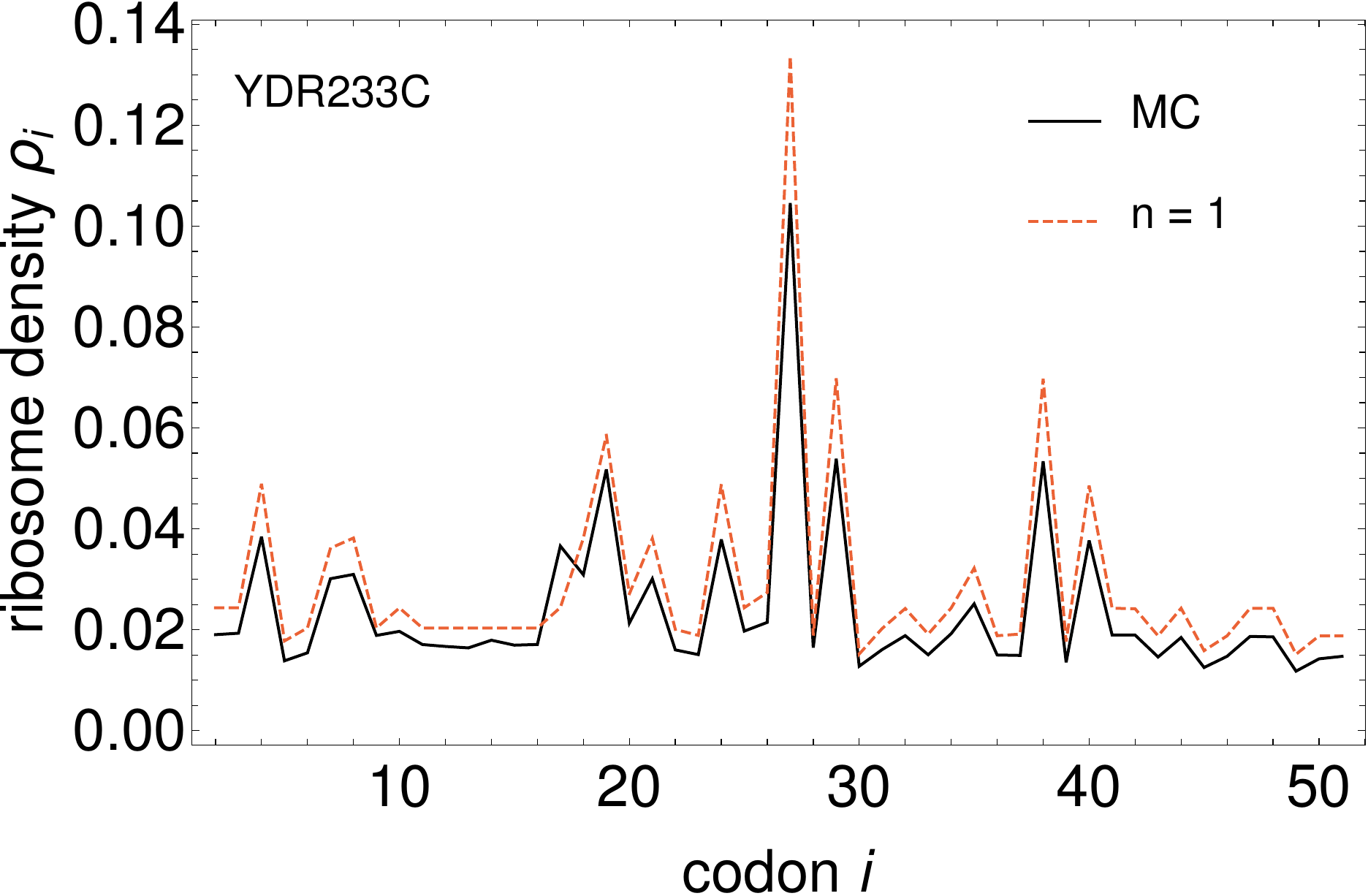}
\includegraphics[width=8cm]{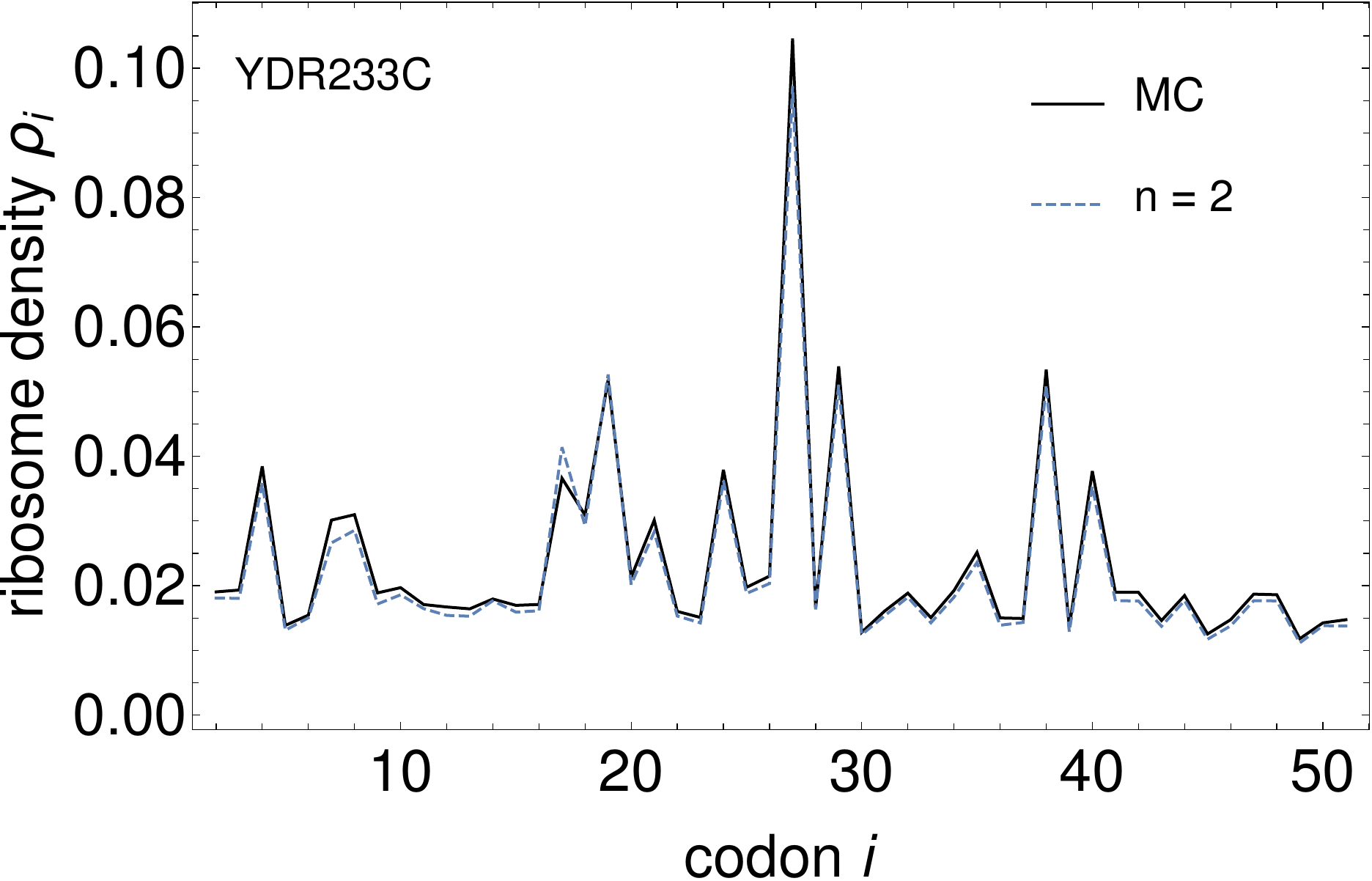}
\caption{Density profiles (first 50 codons) for {\it S. cerevisiae} genes YDR233W and YDR233C. On the left and right are density profiles computed using the first and second order, respectively, and compared to the results of Monte Carlo (MC) simulations. Translation initiation rates are $0.02846$ for YDR223W and $0.21425$ for YDR233C. All results were obtained assuming ribosome drop-off rate $\mu=1.4\cdot 10^{-3}$ s$^{-1}$ and no translation reinitiation ($\gamma=0$).}
\label{fig4}
\end{figure*}
\section{Results}
\subsection{First-order approximation does not account for ribosome interference}

Using (\ref{c0}) and (\ref{c1-i-solution})-(\ref{c1-L+1-solution}) we can compute ribosome density $\rho_i$ and protein synthesis rate $J$ up to the linear order in $\alpha$,
\numparts
\begin{eqnarray}
\label{rho1-i}
&\rho_i\approx\frac{\alpha}{\omega_i}\frac{\prod_{j=2}^{i}\frac{\omega_j}{\omega_j+\mu}}{\left(1-\frac{\gamma}{\beta+\gamma}\prod_{j=2}^{L-1}\frac{\omega_j}{\omega_j+\mu}\right)},\;2\leq i\leq L-1\\
\label{rho1-L}
&\rho_{L}\approx\frac{\alpha}{\beta+\gamma}\frac{\prod_{j=2}^{L-1}\frac{\omega_j}{\omega_j+\mu}}{\left(1-\frac{\gamma}{\beta+\gamma}\prod_{j=2}^{L-1}\frac{\omega_j}{\omega_j+\mu}\right)}\\
\label{J1}
&J\approx\frac{\alpha\prod_{j=2}^{L-1}\frac{\omega_j}{\omega_j+\mu}}{\left(1-\frac{\gamma}{\beta+\gamma}\prod_{j=2}^{L-1}\frac{\omega_j}{\omega_j+\mu}\right)}.
\end{eqnarray}
\endnumparts
These results are similar to the ones obtained by Gilchrist and Wagner using a deterministic model of mRNA translation that includes codon-specific elongation rates, ribosome drop-off and mRNA circularization but ignores ribosome interference \cite{Gilchrist06}. This similarity is not a coincidence but comes from the fact that first order includes configurations with only one ribosome. 

Another interesting prediction from the first order is that the impact of reinitiation strongly depends on the rate of premature termination. That is expected because reinitiation due to mRNA circularisation can only happen if the ribosome has not terminated translation prematurely. The strongest effect is thus when premature termination does not occur, i.e. when $\mu=0$. In that case the products in Eqs. (\ref{rho1-i})-(\ref{J1}) are equal to $1$ and the resulting ribosome density and current read
\numparts
\begin{eqnarray}
\rho_i&\approx&\frac{\alpha(1+\gamma/\beta)}{\omega_i},\;i=2,\dots,L-1\\
\rho_L&\approx&\frac{\alpha}{\beta}\\
J&\approx&\alpha\left(1+\frac{\gamma}{\beta}\right).
\end{eqnarray}
\endnumparts
From here we conclude that in the first-order approximation reinitiation has the same effect as increasing initiation rate from $\alpha$ to $\alpha(1+\gamma/\beta)$.

\subsection{Second-order approximation accounts for ribosome interference}

In the Methods we described in detail how to find all second-order coefficients. This allows us to compute local density $\rho_i$ and current $J$ up to the second order in $\alpha$,
\begin{eqnarray}
\rho_i &=&\rho_{i}^{(1)}\alpha+\rho_{i}^{(2)}\alpha^2\\
J&=&J^{(1)}\alpha+J^{(2)}\alpha^2,
\end{eqnarray}
where linear coefficients $\rho_{i}^{(1)}$ and $J^{(1)}$ are given in Eqs. (\ref{rho1-i}) and (\ref{J1}), respectively, and the second-order coefficients $\rho_{i}^{(2)}$ and $J^{(2)}$ read
\begin{eqnarray}
&\rho_{i}^{(2)}=c_2(1_i)+\sum_{j=2}^{i-\ell}c_2(1_j 1_i)+\sum_{j=i+\ell}^{L}c_2(1_i 1_j)\label{rho2i-2}\\
&J^{(2)}=(\beta+\gamma)c_2(1_L)+\beta\sum_{j=2}^{L-\ell}c_2(1_j 1_L)\nonumber\\
&\quad+\gamma\sum_{j=2+\ell}^{L-\ell}c_2(1_j 1_L)\label{J2-2}.
\end{eqnarray}

Figure \ref{fig4} shows ribosome density (first $50$ codons) for two genes of {\it S. cerevisiae}, YDR223W and YDR233C, computed using the model without reinitiation. These two genes have translation initiation rate smaller than the first quartile and larger than the third quartile of all initiation rates, respectively (see Figure \ref{fig4}). On the left are density profiles computed using the first order and compared with the results of Monte Carlo simulations. As expected, the agreement is worse for the gene that has a larger value of $\alpha$. On the right are density profiles obtained using the second order, which agree well with the results of Monte Carlo simulations.

\begin{figure*}[htb]
\centering
\includegraphics[width=8cm]{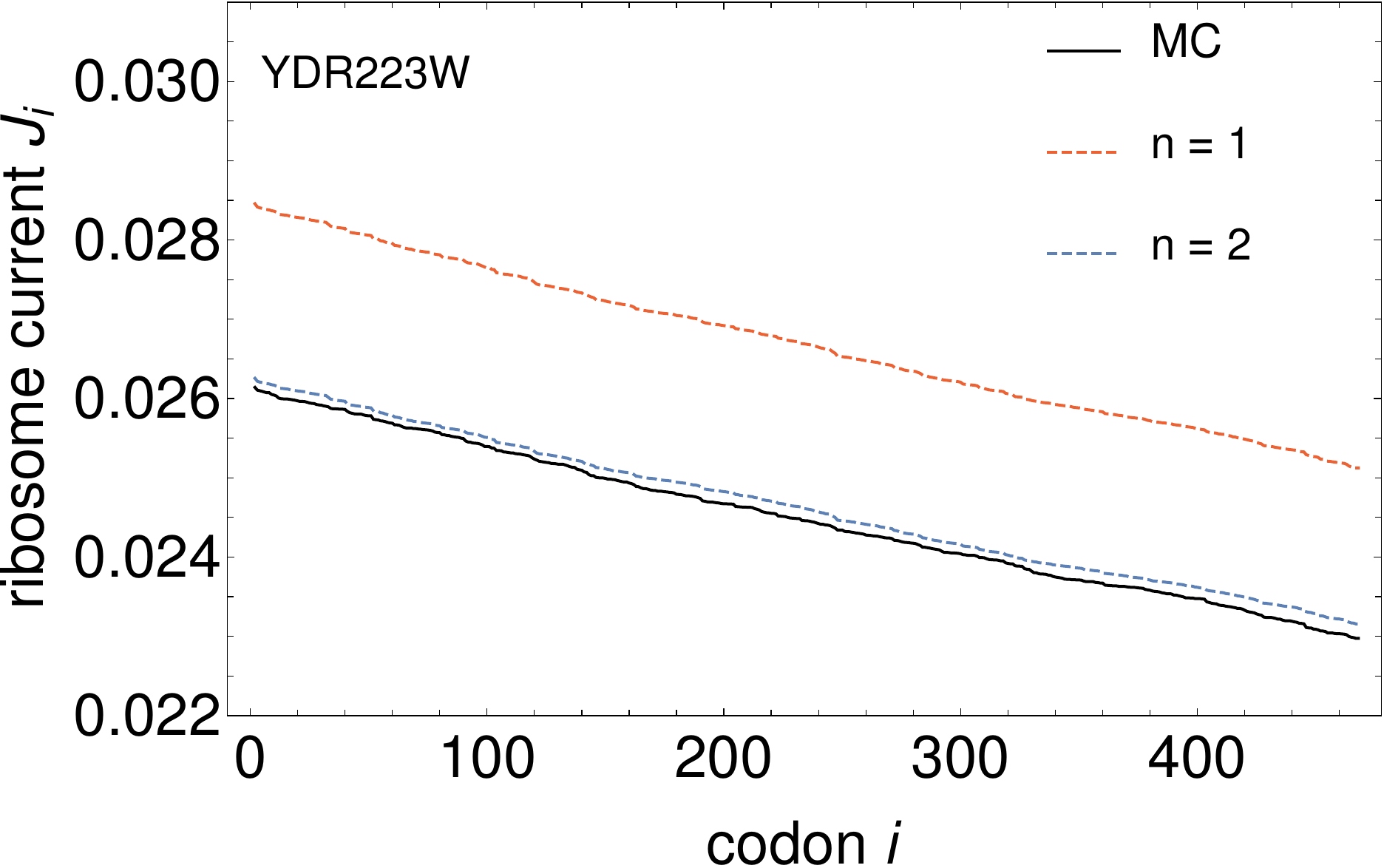}
\includegraphics[width=8cm]{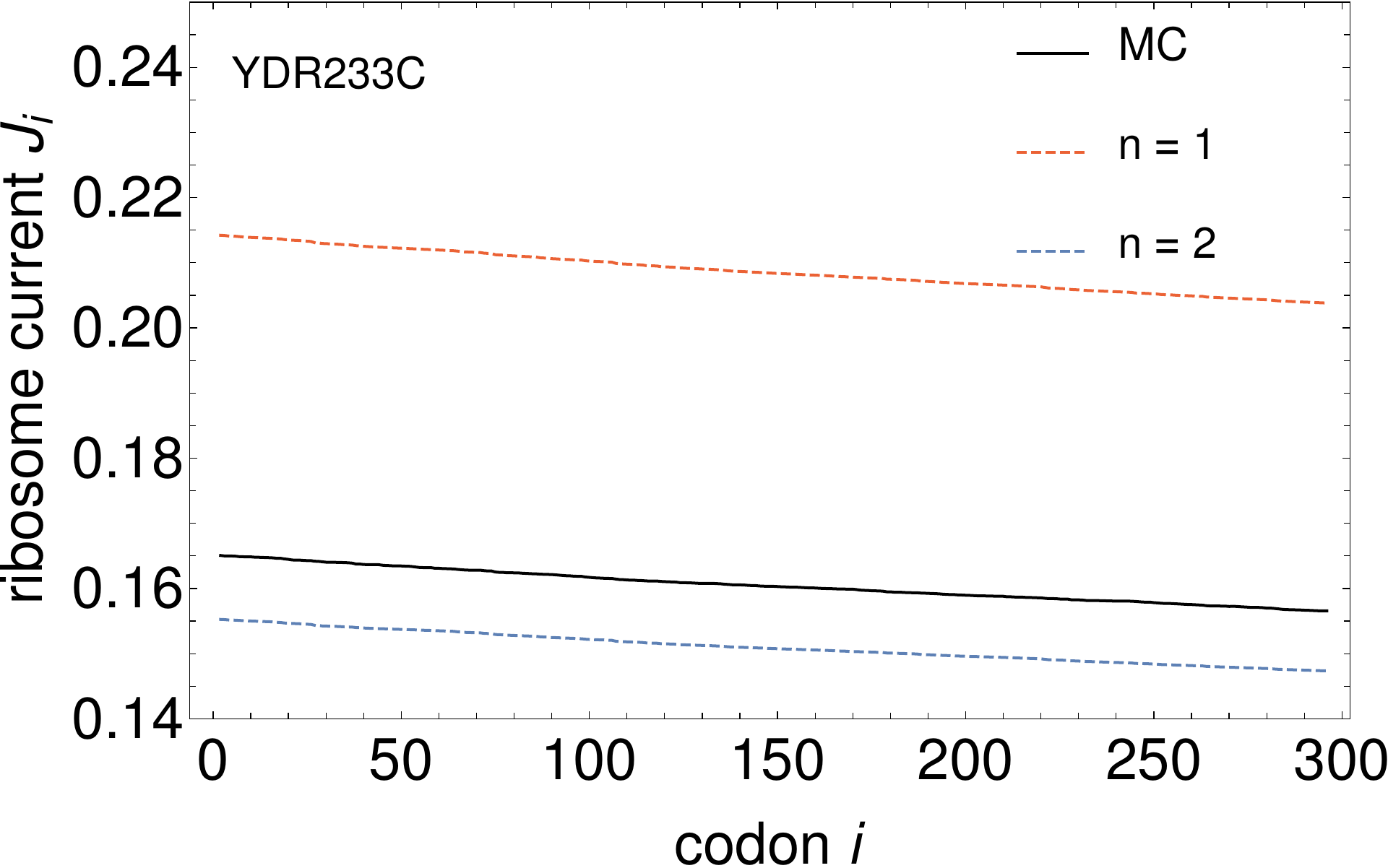}\\
\caption{Ribosome current $J_i$ across the mRNA for {\it S. cerevisiae} genes YDR233W and YDR233C, computed from Eq. (\ref{current_i}). Solid black line is the result of stochastic simulations, while red and blue dashed lines represent  first-order and second-order approximation, respectively. All results were obtained assuming ribosome drop-off rate $\mu=1.4\cdot 10^{-3}$ s$^{-1}$ and no translation reinitiation ($\gamma=0$).}
\label{fig5}
\end{figure*}

In Figure \ref{fig5} we show ribosome current $J_i$ across the mRNA, computed from Eq. (\ref{current_i}) for the same two genes as before and using the model without reinitiation. Unlike the density, the first-order approximation of the current already shows a significant discrepancy compared to Monte Carlo simulations for both genes. As expected, the discrepancy is reduced when using second-order approximation. 
\subsection{Effect of ribosome interference on second-order coefficients}

Because the second order must be computed numerically, how exactly the second-order coefficients are affected by ribosome interference is not immediately obvious. If we imagine a mathematical model in which ribosome interference is ignored, we would expect $P(C)$ to be a product of single-particle weights $c_1(1_i)\alpha$
\begin{eqnarray}
P(C)&=&\frac{1}{Z_L}\prod_{j=1}^{N(C)}\alpha c_1(1_{X(j)})\nonumber\\
&=&\frac{1}{Z_L}\prod_{i=2}^{L}\left[\tau_i c_1(1_i)\alpha+(1-\tau_i)\right],
\end{eqnarray}
where $N(C)$ is the number of particles in $C$, $X(j)$ is the position of the $j$-th particle and $Z_L=\prod_{i=2}^{L}(1+c_1(1_i)\alpha)$ is the normalization (see Ref. \cite{Nossan18} for more details in which we termed this approximation the independent particle approximation or IPA). Taking $C=1_i 1_j$ and expanding $P(C)$ in $\alpha$ up to the quadratic order we get
\begin{equation}
c_2(1_i 1_j)\stackrel{\textrm{IPA}}{=}c_1(1_i)c_1(1_j).
\end{equation}

Going back to the model with exclusion, we can write $c_2(1_i 1_j)$ as
\begin{equation}
\label{g2ij}
c_2(1_i 1_j)=c_1(1_i)c_1(1_j)g_2(1_i 1_j).
\end{equation}
where $g(1_i 1_j)$ measures the deviation from the IPA (for which $g(1_i 1_j)=1$), i.e. the effect of exclusion. The equations for $g_2(1_i 1_j)$ for $i\neq 2$ and $j\neq L$ read
\numparts
\begin{eqnarray}
\label{g2iil-equation}
&g_2(1_i 1_{i+\ell})=\frac{e(1_i)}{e(1_{i+\ell})}g_2(1_{i-1}1_{i+\ell}),\;i\neq 2\\
\label{g2ij-equation}
&g_2(1_i 1_j)=\frac{e(1_i)}{e(1_i)+e(1_j)}g_2(1_{i-1}1_j)\nonumber\\
&\quad+\frac{e(1_j)}{e(1_i)+e(1_j)}g_2(1_i 1_{j-1}),\; i\neq 2,j\neq L,
\end{eqnarray}
\endnumparts
where $e(1_i)=(1-\delta_{i,L})(\omega_i+\mu)+\delta_{i,L}\beta$. We notice that Eq. (\ref{g2ij-equation}) could be solved by setting all $g_2$ to $1$, however that would violate the initial equation (\ref{g2iil-equation}). On the other hand, both $e(1_i)/(e(1_i)+e(1_j))$ and $e(1_j)/(e(1_i)+e(1_j))$ in Eq. (\ref{g2ij}) are strictly less than $1$, which means that any deviation of $g_2$ from $1$ in Eq. (\ref{g2iil-equation}) will be attenuated by subsequent iterations of Eq. (\ref{g2ij-equation}). Therefore we expect to find $g_2(1_i 1_j)\approx 1$ when codons $i$ and $j$ are far apart, i.e.
\begin{equation}
c_2(1_i 1_j)\approx c_1(1_i)c_1(1_j)\quad\textrm{for $\vert i-j\vert\gg\ell$}.
\end{equation}
Certainly, the effect of exclusion is strongest when the ribosomes are next to each other, i.e. for $j=i+\ell$. In that case there is either a magnification ($e(1_i)>e(1_{i+\ell})$) or reduction ($e(1_i)<e(1_{i+\ell})$) in $g_2(1_i 1_j)$ compared to the IPA that is carried over to the surrounding codons. 

\begin{figure}[hbt]
\centering\includegraphics[width=8cm]{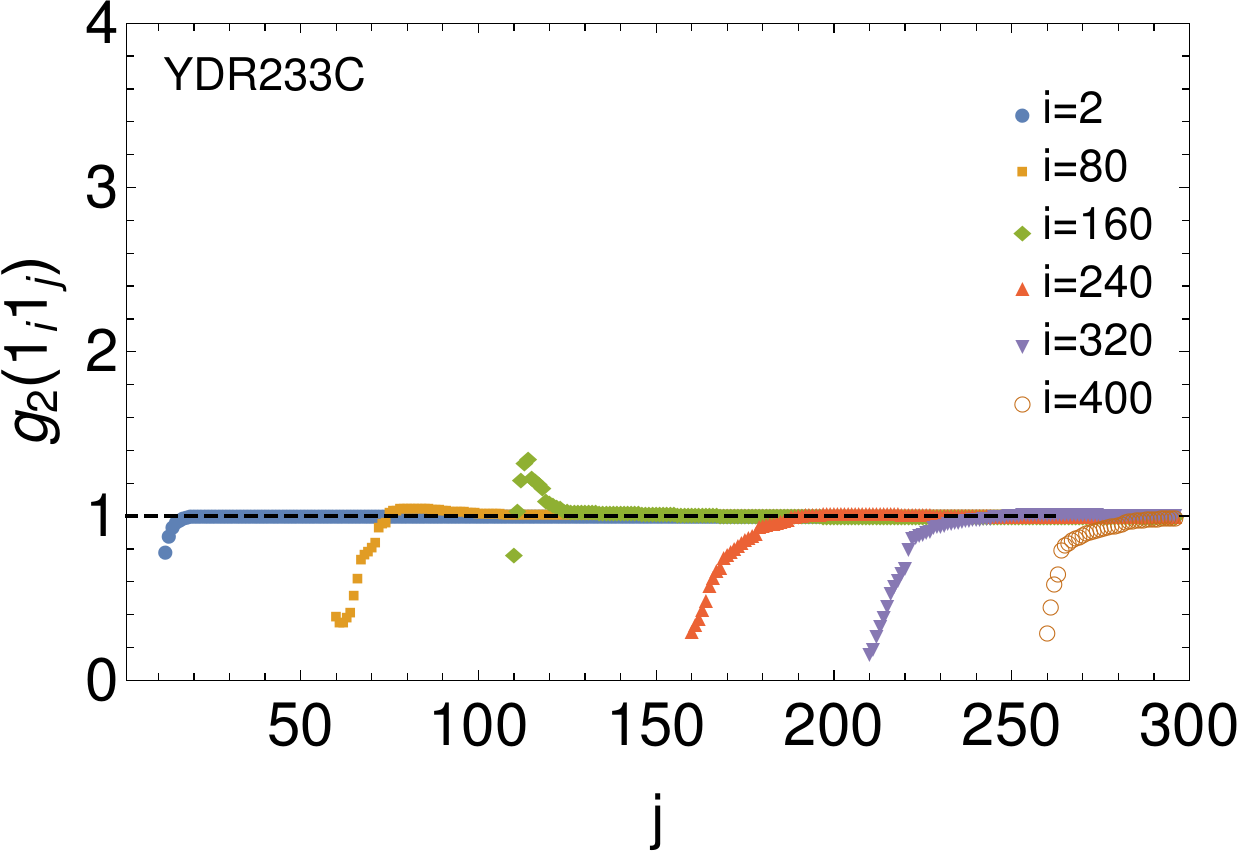}
\caption{The coefficient $g_2(1_i 1)j)$ for YDR233C as a function of $j$ for several values of $i$ and assuming no translation reinitiation.}
\label{fig6}
\end{figure}

In Figure \ref{fig6} we plot $g_2(1_i 1_j)$ for YDR233C gene as a function of $j$ for several values of $i$. As predicted, the deviation of $g_2(1_i 1_j)$ from $1$ is the largest at $j=i+\ell$ and eventually decays to $1$ as $j$ gets away from $i$.

\subsection{High-order approximations are needed for genes with high initiation rates}

\begin{figure*}[htb]
	\centering
	\includegraphics[width=8cm]{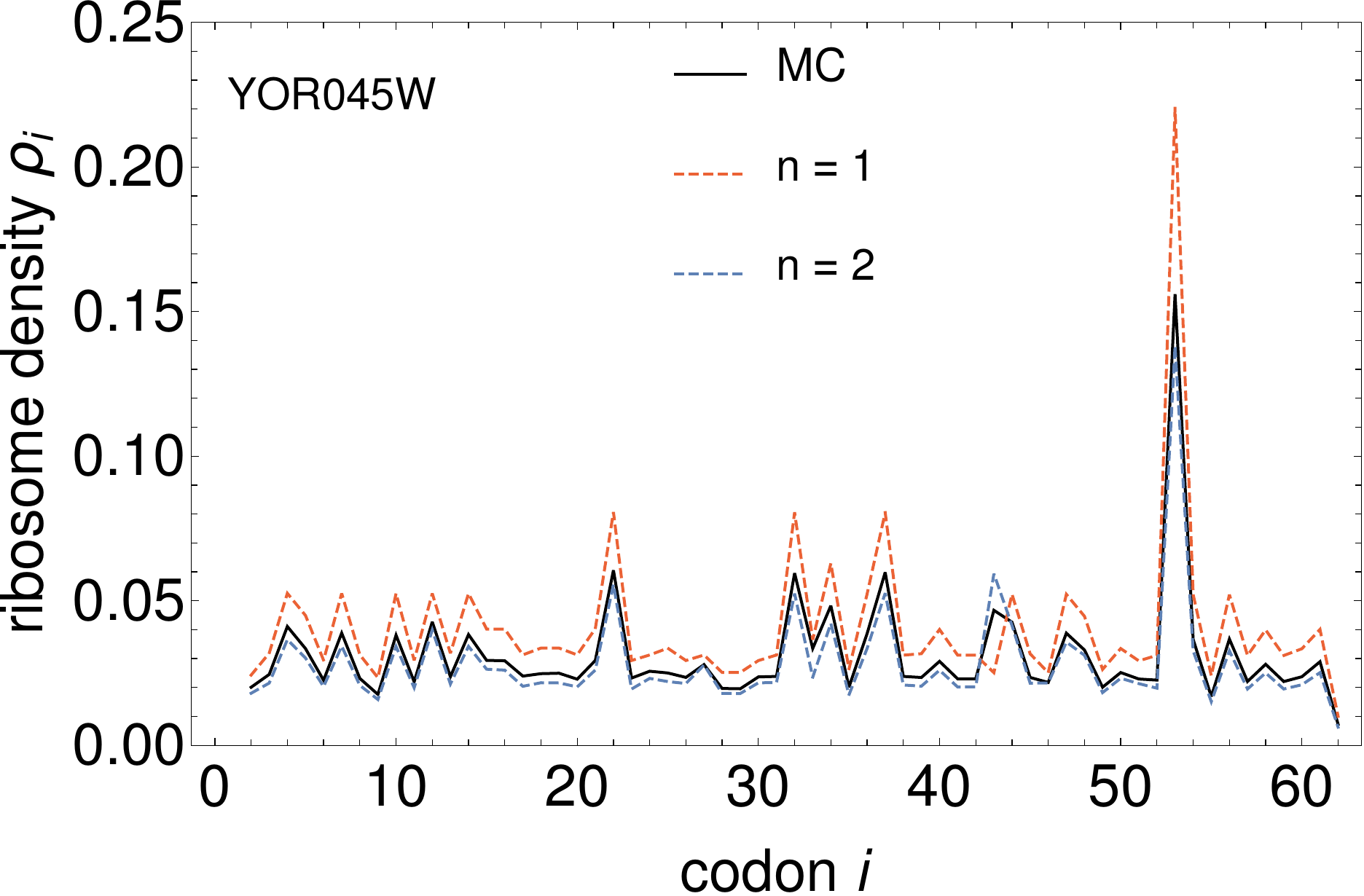}
	\includegraphics[width=8cm]{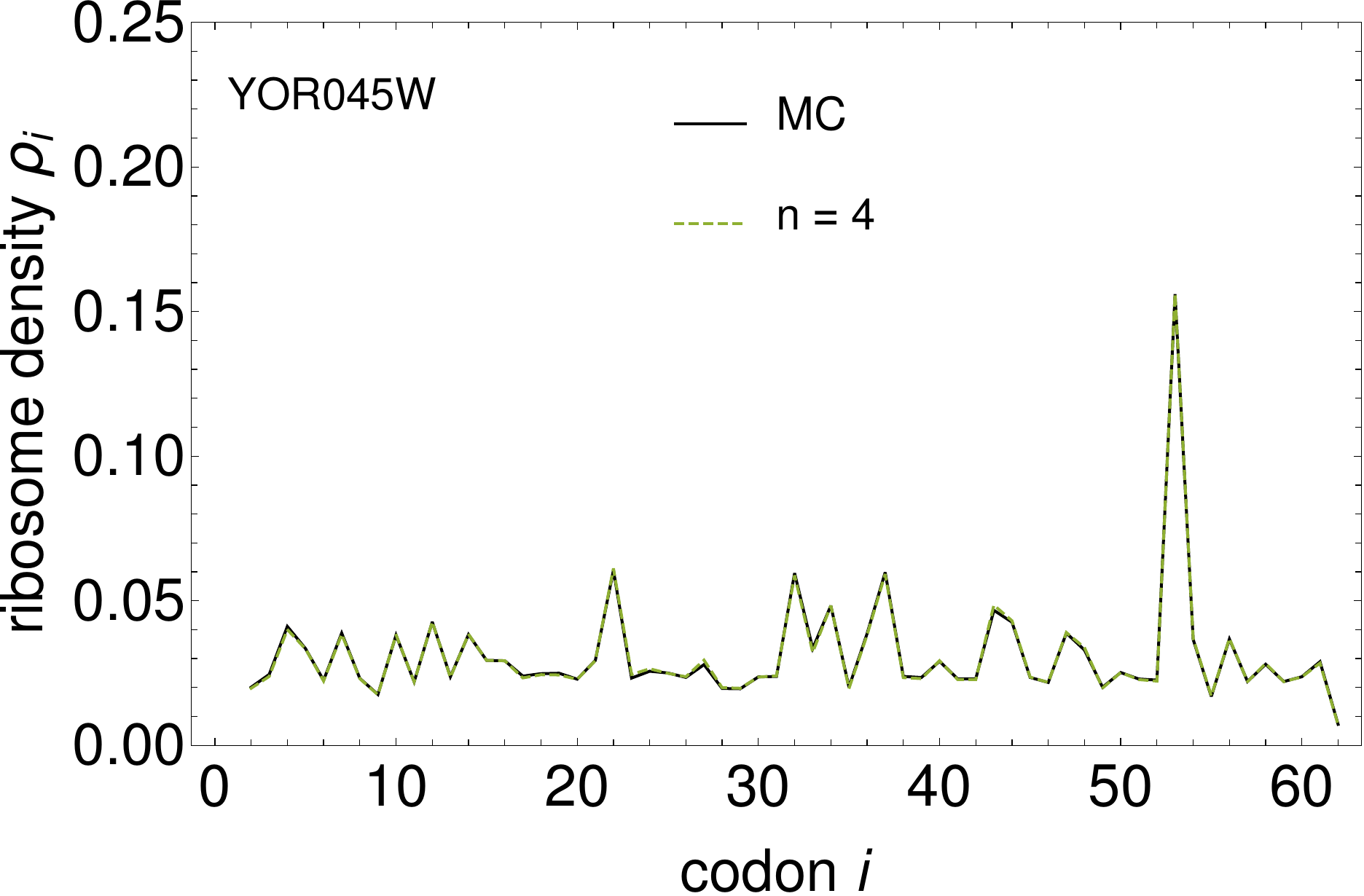}
	\caption{Ribosome density profiles for {\it S. cerevisiae} gene YOR045W. On the left and right are density profiles computed using the second and fourth order, respectively, and compared to the results of Monte Carlo (MC) simulations. Translation initiation rate is $0.35423$. All results were obtained assuming ribosome drop-off rate $\mu=1.4\cdot 10^{-3}$ s$^{-1}$ and no translation reinitiation ($\gamma=0$).}
	\label{fig7}
\end{figure*}

\begin{figure}[hbt]
	\centering
	\includegraphics[width=8cm]{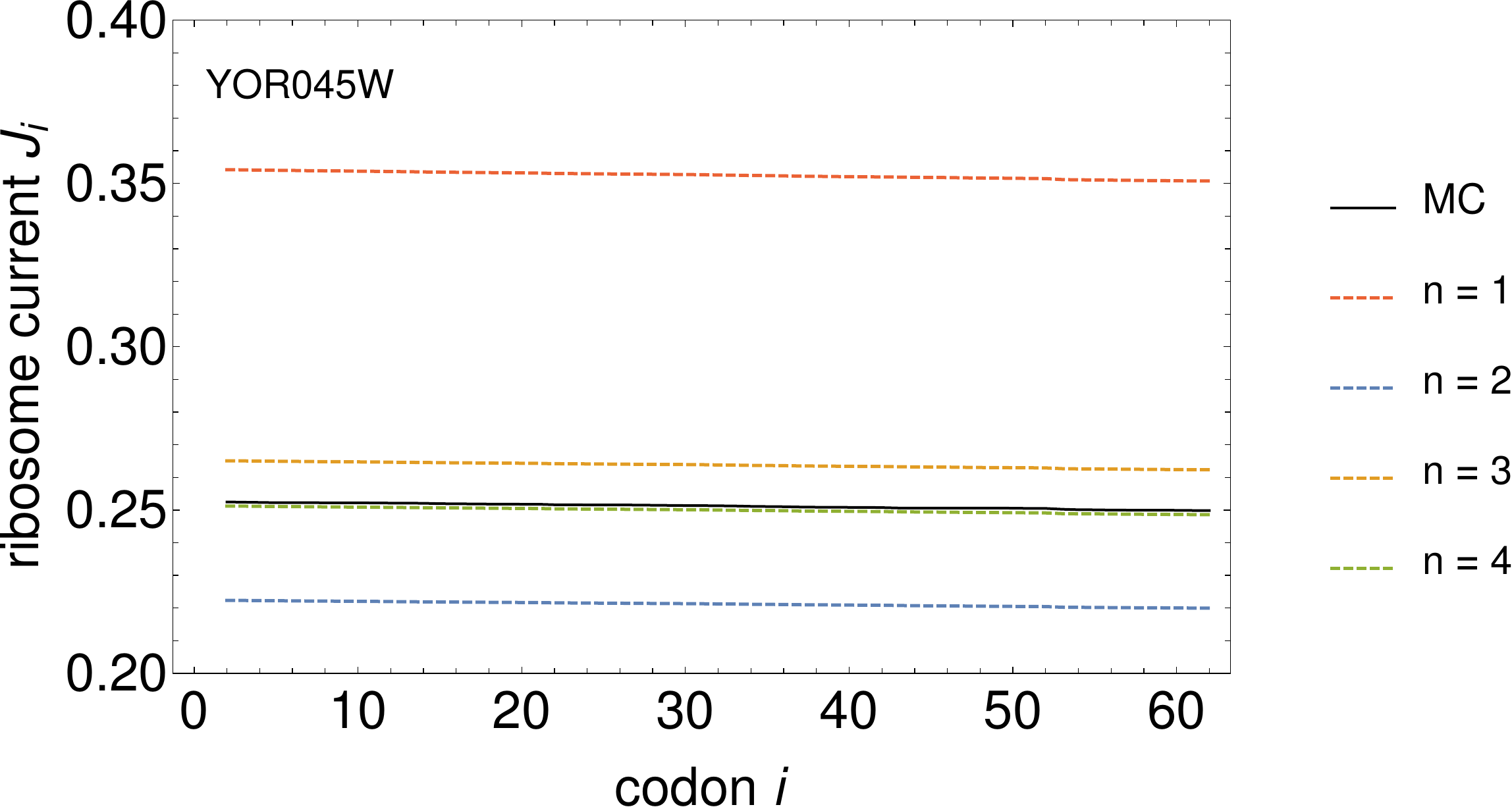}
	\caption{Ribosome current $J_i$ across the mRNA for {\it S. cerevisiae} gene YOR045W, computed from Eq. (\ref{current_i}). Solid black line is the result of stochastic simulations, while red, blue, orange and green dashed lines represent first-order, second-order, third-order and fourth-order approximation, respectively. All results were obtained assuming ribosome drop-off rate $\mu=1.4\cdot 10^{-3}$ s$^{-1}$ and no translation reinitiation ($\gamma=0$).}
	\label{fig8}
\end{figure}

As the rate of initiation increases, using the first-order or second-order approximation may lead to significant errors. In Figure \ref{fig7} we demonstrate this for gene YOR045W, which has a relatively large value of $\alpha=0.35423$ and total ribosome density $\rho=0.03256$ (approximately $33$\% of the maximum theoretical density $1/\ell=0.1$). On the left are density profiles computed using first-order and second-order approximation and compared to the results of Monte Carlo simulations. On the right is the density profile obtained using the fourth-order approximation, which agrees well with the results of Monte Carlo simulations. Similar conclusions can be made for the ribosome current $J_i$ across the mRNA., see Figure \ref{fig8}.

\subsection{Translation reinitiation has the same effect as increasing initiation rate}

\begin{figure*}[htb]
\centering
\includegraphics[width=8cm]{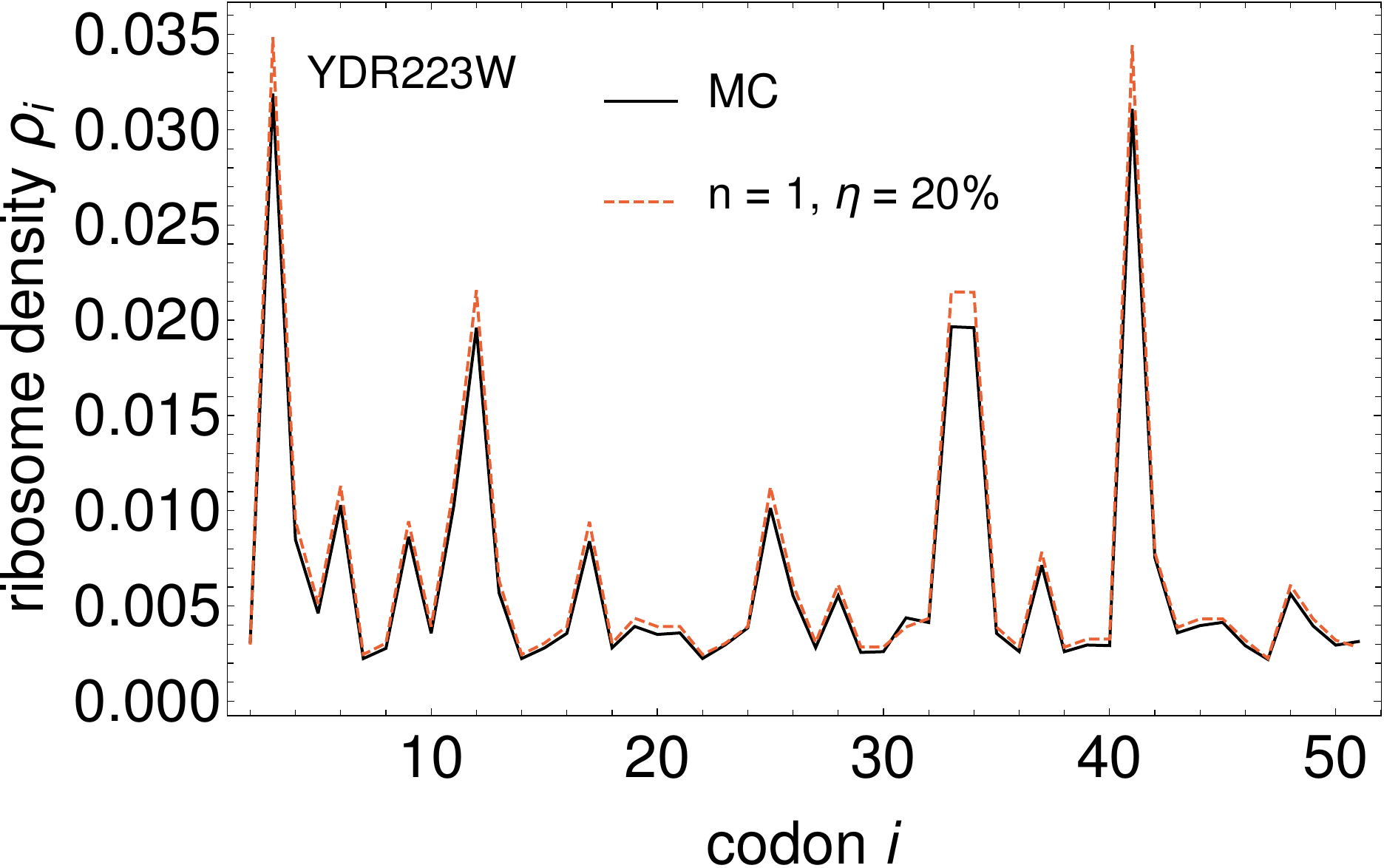}
\includegraphics[width=8cm]{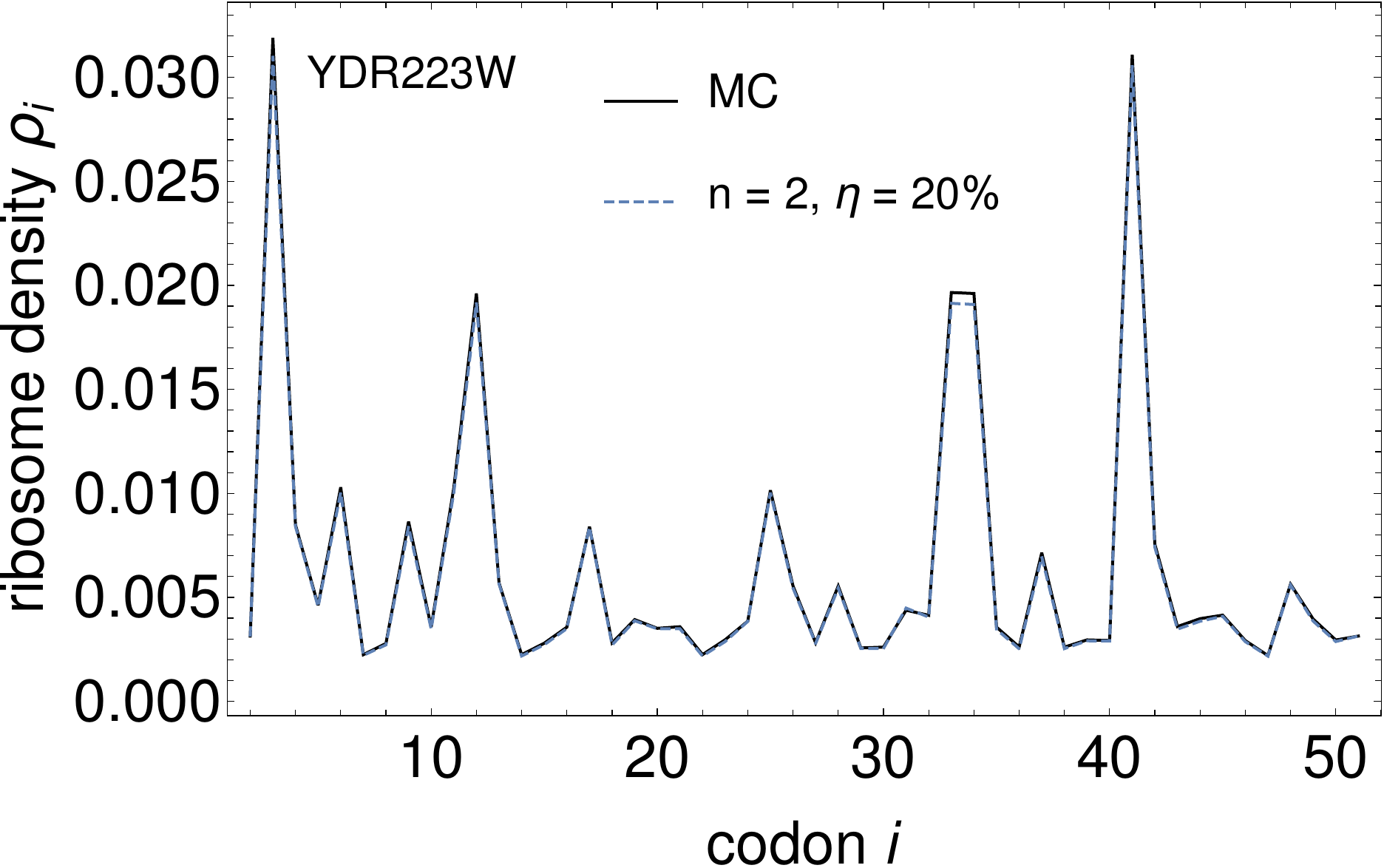}
\vspace{1em}
\includegraphics[width=8cm]{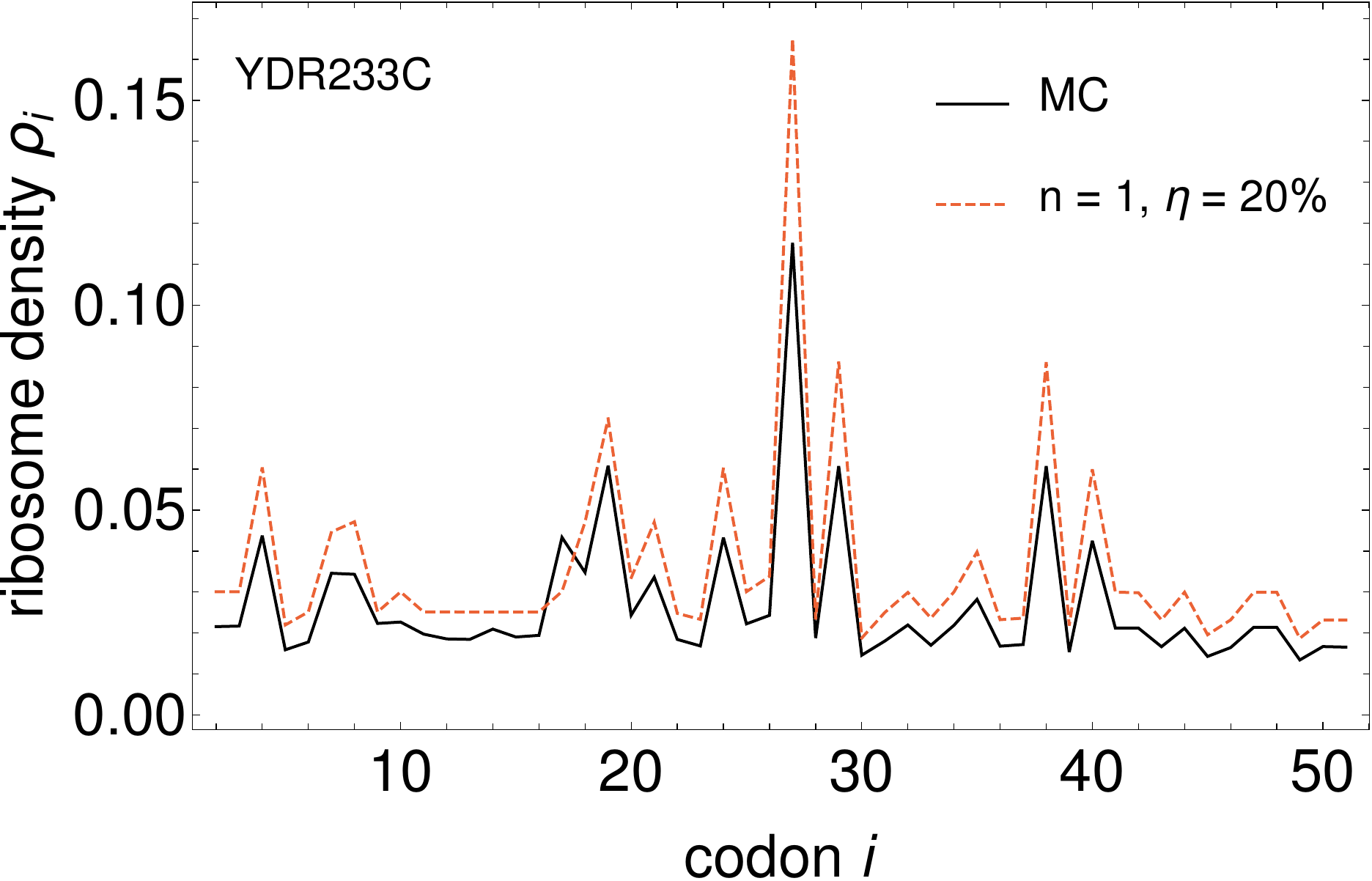}
\includegraphics[width=8cm]{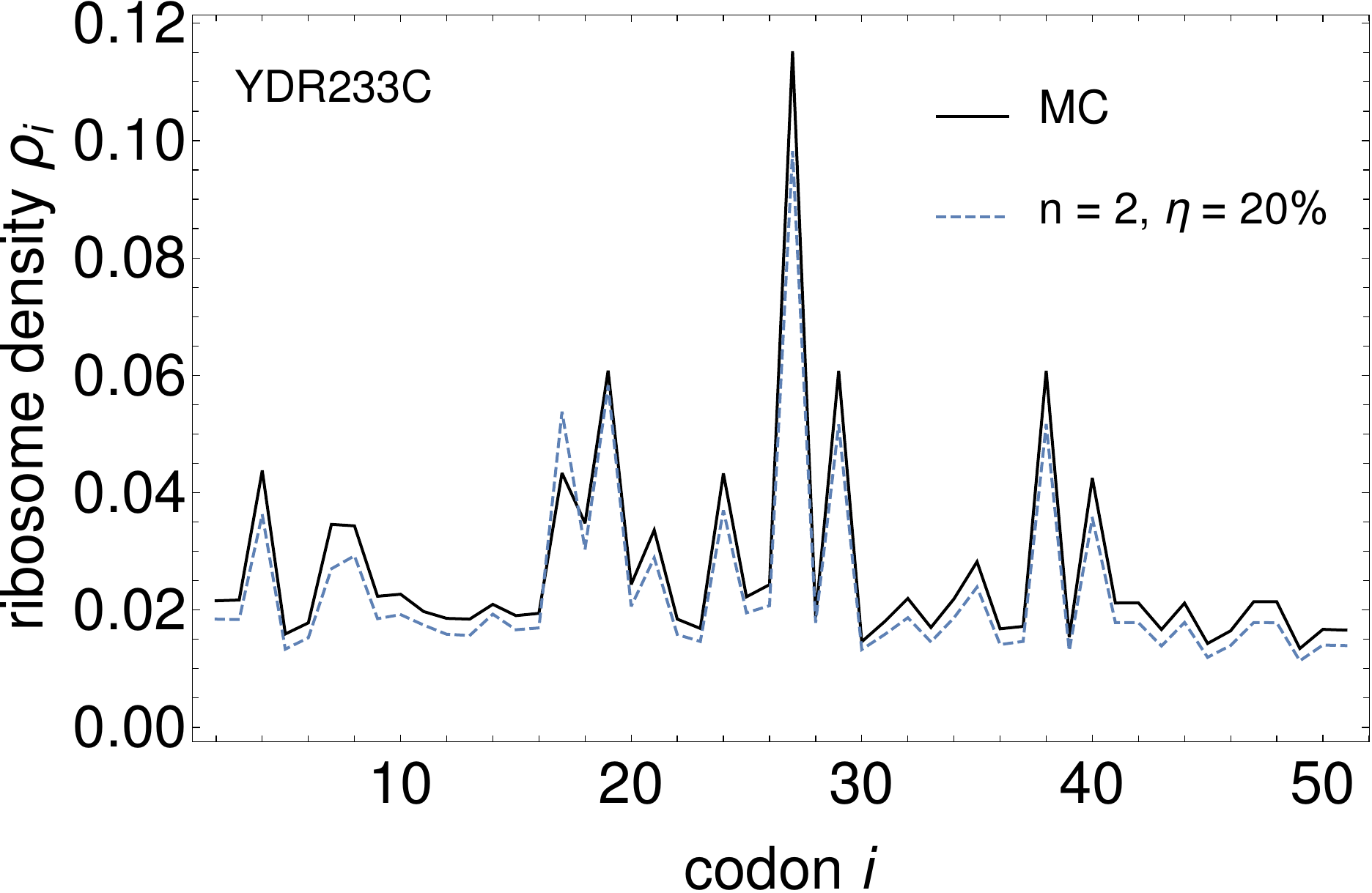}
\caption{Density profiles (first 50 codons) for {\it S. cerevisiae} genes YDR233W and YDR233C. On the left and right are density profiles computed using the first and second order, respectively, and compared to the results of Monte Carlo (MC) simulations. Translation initiation rates are $0.02846$ for YDR223W and $0.21425$ for YDR233C. All results were obtained assuming ribosome drop-off rate $\mu=1.4\cdot 10^{-3}$ s$^{-1}$ and translation reinitiation with $\eta=0.2$.}
\label{fig9}
\end{figure*}

In Figure \ref{fig9} we present density profiles for two genes, YDR223W and YDR233C, obtained using a model with translation reinitiation with reinitiation efficiency set to $\eta=0.2$.

For gene YDR223W, which has a small value of $\alpha$, the agreement between the second-order approximation and results of Monte Carlo simulations is excellent. On the other hand, there is a visible discrepancy between the second-order approximation and results of Monte Carlo simulations for gene YDR233C, which has a relatively large value of $\alpha$. This result is expected because translation reinitiation increases the number of ribosomes that initiate translation, which in turn may require more terms in the series expansion. Therein lies the problem--computing higher-order terms in the model with translation reinitiation is not as straightforward as without reinitiation, because it involves solving a linear system of equations. 

\begin{figure*}[htb]
\centering
\includegraphics[width=8cm]{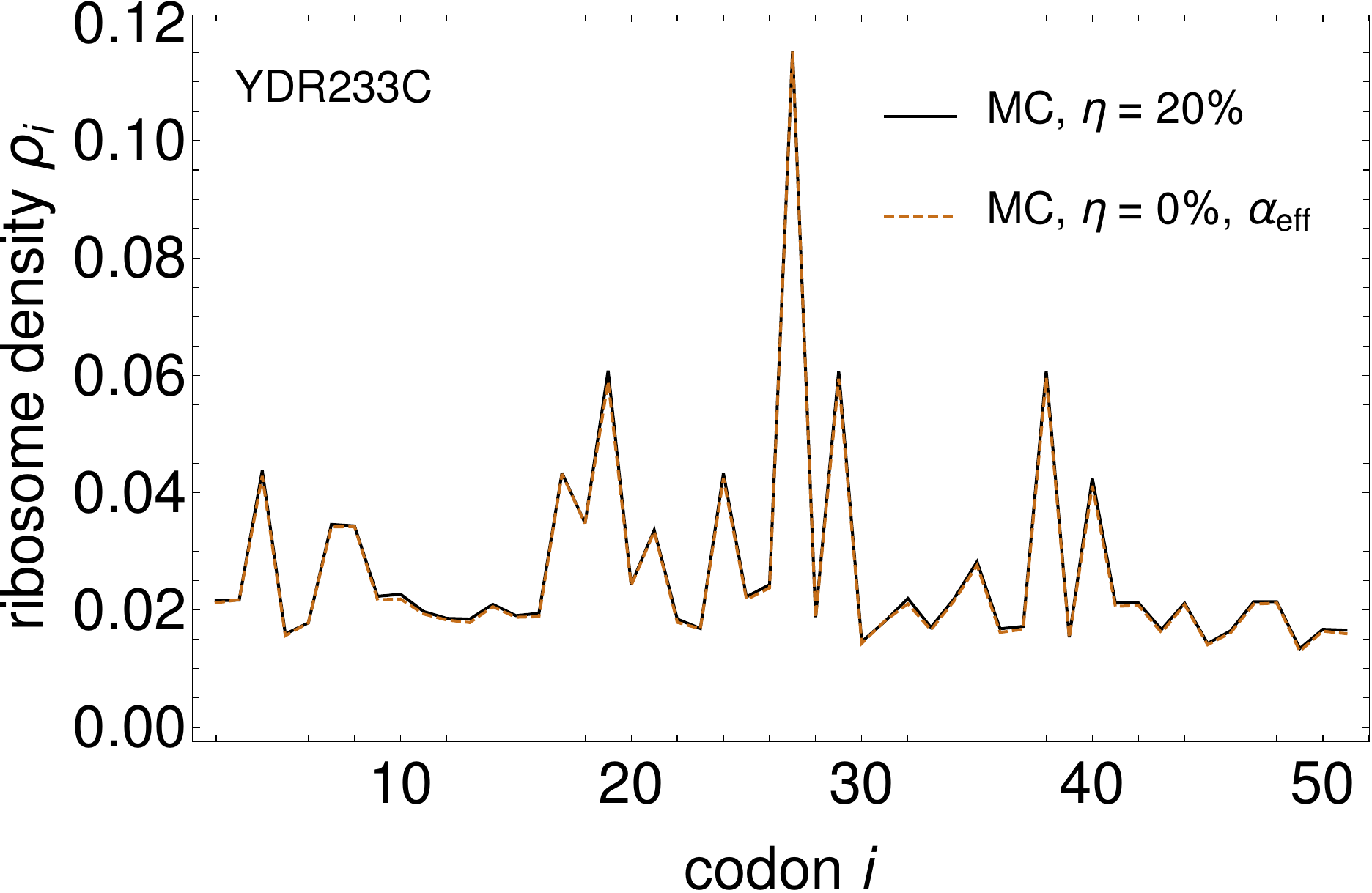}
\includegraphics[width=8cm]{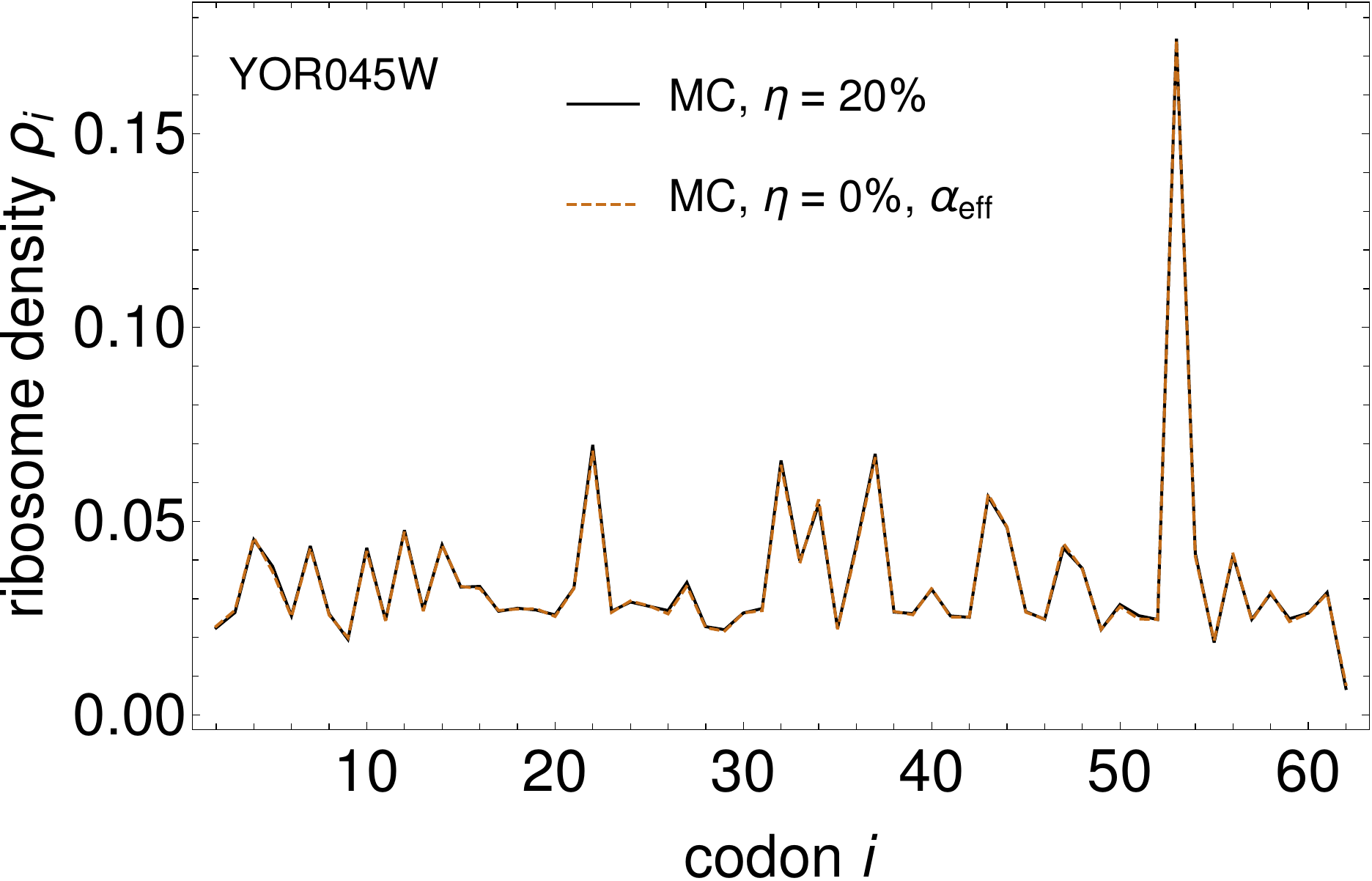}
\caption{Density profiles for {\it S. cerevisiae} genes YDR233W (first 50 codons) and YDR233C (all codons), obtained using Monte Carlo simulations of the model with reinitiation ($\eta=0.2$) and the effective model without reinitiation ($\eta=0$). Translation initiation rates are $\alpha=0.21425$ for YDR233C and $0.35423$ for YOR045W. All results were obtained assuming ribosome drop-off rate $\mu=1.4\cdot 10^{-3}$ s$^{-1}$.}
\label{fig10}
\end{figure*}

Here we take a pragmatic approach to tackle this problem. We ask if the model with translation reinitiation can be replaced with an effective model without reinitiation but in which the rate of translation initiation is set to a higher value $\alpha_{\textrm{eff}}>\alpha$. This value must be such that both models yield the same predictions for the ribosome density $\rho_i$ and current $J$. The way to achieve this is to set 
\begin{eqnarray}
    \alpha_{\textrm{eff}}&=&\frac{J_{\textrm{in}}}{\left\langle\prod_{i=2}^{\ell+1}(1-\tau_i)\right\rangle}=\alpha+\gamma\frac{\left\langle\tau_L\prod_{i=2}^{\ell+1}(1-\tau_i)\right\rangle}{\left\langle\prod_{i=2}^{\ell+1}(1-\tau_i)\right\rangle}\nonumber\\
    &=&\alpha+\gamma\frac{J-\beta\langle\tau_L\rangle}{\left\langle\prod_{i=2}^{\ell+1}(1-\tau_i)\right\rangle}\label{alpha-eff}
\end{eqnarray}
where $J_{\textrm{in}}$ is the total influx of ribosomes initiating translation, Eq. (\ref{initial_current}), and the denominator is the probability that the first $\ell=10$ codons are not occupied by another ribosome's A-site. In Figure \ref{fig10} we present density profiles for genes YDR233C and YOR045W obtained using Monte Carlo simulations of the model with reinitiation and the effective model without reinitiation. For both genes we find an excellent agreement between the two models.

This result has two important implications. The first one is technical--we can apply the power series method to the effective model and avoid the problem of solving a linear system of equations. The second one is biological. If we want to estimate the rate of initiation $\alpha$ by matching theoretical density $\rho(\alpha)$ to the experimental density from polysome profiling experiments, as it was done in Ref. \cite{Ciandrini13}, we cannot truly distinguish reinitiation from {\it de nuovo} initiation. In other words, the evidence for translation reinitiation may be very difficult to find experimentally because the effect of translation reinitiation is the same as {\it de novo} initiation at a higher rate.

\section{Discussion}

Our first main result is that the power series method is applicable to the TASEP with ribosome drop-off and translation reinitiation. This complements previous work in which the method was applied to the TASEP with multi-step elongation \cite{Nossan18}. 
We tested the method on {\it Saccharomyces cerevisiae} under physiological conditions and found that the model-predicted ribosome density and current are faithfully described by the second-order approximation for most of the genes. Interestingly, second order is the lowest order at which ribosome interference occurs, suggesting that ribosome interference does have an effect on translation. This is clearly visible for genes with high initiation rates belonging to the last quartile in Figure \ref{fig3}, for which higher-order approximations are needed to describe the data. In that sense the statement often found in biology that initiation is rate-limiting for translation is true \cite{Shah13}, but incomplete--translation elongation does have an effect on translation.

Our second main result is an iterative algorithm that computes ribosome density and current up to any order. This is a significant improvement over previous work that considered only second order \cite{Nossan18}. At the moment computing orders beyond the second is limited to the model without translation reinitiation. The problem is that reinitiation does not allow for the coefficients $c_n(C)$ in Eq. (\ref{power-series}) to be found recursively starting from a configuration with all ribosomes stacked to the left. Instead one must first solve a closed linear system of equations for the coefficients $c_n(C)$ with $n$-th ribosome at the last codon site (the stop codon). We believe this technical issue will be resolved in the future. More serious limitation is that the number of configurations contributing to $n$-th order is of order of $L^n$. This is a problem because the coefficients are computed recursively and need to be stored during the recursion process, which limits how large $n$ and $L$ can be.

TASEP-based models of translation are usually studied using approximations (called mean-field approximations) that ignore correlations between neighbouring ribosomes \cite{MacDonald68,MacDonald69,Shaw04}. Power series method is the only method available that can account for these correlations. In this work we studied the effect of ribosome-ribosome correlations on the second-order coefficients $c_2(1_i 1_j)$ for the TASEP without translation reinitiation. The strongest correlations were found for ribosomes that are next to each other ($j=i+\ell$), with the strength of correlations depending on the ratio $(\omega_i+\mu)/(\omega_{i+\ell}+\mu)$. For $(\omega_i+\mu)/(\omega_{i+\ell}+\mu)<1$ ($(\omega_i+\mu)/(\omega_{i+\ell}+\mu)>1$), the density at codon $i$ is smaller (larger) than it would be on a mRNA composed of only one ribosome. Taking this further, if we could arrange codons in a sequence such that
\begin{equation}
    \omega_2<\omega_3<\dots<\omega_L,
    \label{ramp}
\end{equation}
then according to the second-order approximation, the total ribosome density for that sequence would be minimal compared to the same choice of codons arranged in a different sequence. This is an interesting result when put in the context of {\it ramp hypothesis} proposed by Tuller \etal \cite{Tuller10}, who found that the first 30--50 codons are, on average, translated at slow elongation speeds. The ramp hypothesis states that slow elongation speeds at the beginning reduce ribosome traffic jams and thus have a purpose of minimising the cost of protein production. Our hypothetical arrangement in Eq. (\ref{ramp}), which could be considered as a perfect ramp, is unlikely to occur in real codon sequences due to other evolutionary factors driving codon usage. Nevertheless, our findings may provide the first step in understanding the origin of the ramp from a mathematical point of view.

\section{Conclusions}

We have presented a versatile method for studying TASEP-based models that account for several mechanistic details of the translation process: codon-dependent elongation, premature termination and mRNA circularisation. We have applied our method to the model organism {\it Saccharomyces cerevisiae} under physiological conditions and found an excellent agreement for the ribosome density and current with the results of stochastic simulations.

While the TASEP as a model for translation has been proposed half a century ago, it has only recently become common in computational biology. Our goal for the future is to use the presented method for analysing biological data e.g. from ribosome profiling experiments, which would give us a better understanding of the translation process and allow us to address open questions in the cell biology.

\ack
JSN was supported by the Leverhulme Trust Early Career Fellowship under grant number ECF-2016-768.

\end{document}